\documentclass[aps,eqsecnum,preprint,floats,epsf,epsfig,nofootinbib,letter]{revtex4}
\usepackage{epsfig}

\begin{document}
\def\be{\begin{eqnarray}}
\def\en{\end{eqnarray}}
\def\non{\nonumber}
\def\la{\langle}
\def\ra{\rangle}
\def\nc{N_c^{\rm eff}}
\def\vp{\varepsilon}
\def\drho{\bar\rho}
\def\deta{\bar\eta}
\def\CP{{\it CP}~}
\def\a{{\cal A}}
\def\B{{\cal B}}
\def\c{{\cal C}}
\def\d{{\cal D}}
\def\e{{\cal E}}
\def\p{{\cal P}}
\def\t{{\cal T}}
\def\up{\uparrow}
\def\dw{\downarrow}
\def\vma{{_{V-A}}}
\def\vpa{{_{V+A}}}
\def\smp{{_{S-P}}}
\def\spp{{_{S+P}}}
\def\J{{J/\psi}}
\def\ov{\overline}
\def\Lqcd{{\Lambda_{\rm QCD}}}
\def\pr{{Phys. Rev.}~}
\def\prl{{Phys. Rev. Lett.}~}
\def\pl{{Phys. Lett.}~}
\def\np{{Nucl. Phys.}~}
\def\zp{{Z. Phys.}~}
\def\lsim{ {\ \lower-1.2pt\vbox{\hbox{\rlap{$<$}\lower5pt\vbox{\hbox{$\sim$}
}}}\ } }
\def\gsim{ {\ \lower-1.2pt\vbox{\hbox{\rlap{$>$}\lower5pt\vbox{\hbox{$\sim$}
}}}\ } }

\font\el=cmbx10 scaled \magstep2{\obeylines\hfill February, 2006}

\vskip 1.5 cm

\centerline{\large\bf Exclusive Baryonic $B$ Decays Circa 2005}
\bigskip
\centerline{\bf Hai-Yang Cheng}
\medskip

\medskip
\centerline{Institute of Physics, Academia Sinica}
\centerline{Taipei, Taiwan 115, Republic of China}

\medskip
\bigskip
\bigskip
\centerline{\bf Abstract}
\bigskip

\small The status of exclusive two-body and three-body baryonic
$B$ decays is reviewed. The threshold enhancement effect in the
dibaryon invariant mass and the angular distributions in the
dibaryon rest frame are stressed and explained. Weak radiative
baryonic $B$ decays mediated by the electromagnetic penguin
process $b\to s\gamma$ are discussed. Puzzles with the correlation
observed in $B^-\to p\bar pK^-$ decay and the unexpectedly large
rate observed for $B\to\Lambda_c\bar\Lambda_cK$ are examined. The
former may indicate that the $p\bar p$ system is produced through
some intermediate states, while the latter implies the failure of
naive factorization for $\Lambda\bar\Lambda K$ modes and may hint
at the importance of final-state rescattering effects.

\pagebreak

\small

\section{Introduction}

A unique feature of hadronic $B$ decays is that the $B$ meson is
heavy enough to allow a baryon-antibaryon pair production in the
final state.\footnote{In charm decay, $D_s^+\to p\bar n$ is the
only baryonic $D$ decay mode which is physically allowed. However,
its branching ratio is expected to be very small, of order
$10^{-6}$, because of partial conservation of axial current
\cite{Pham}.}
During the Lepton-Photon Conference in 1987, ARGUS announced the
first measurement of the decay modes $p\bar p\pi^\pm$ and $p\bar
p\pi^+\pi^-$ in $B$ decays at the level of $10^{-4}$ \cite{ARGUS}.
Although this evidence of charmless baryonic $B$ decays was
immediately ruled out by CLEO \cite{CLEO89}, it nevertheless has
stimulated extensive theoretical studies during the period of
1988-1992. Several different model frameworks have been proposed:
the constituent quark model \cite{Korner}, the pole model
\cite{DTS,Jarfi}, the QCD sum rule \cite{Chernyak}, the diquark
model \cite{Ball} and flavor symmetry considerations
\cite{Gronau}.

However, experimental and theoretical activities towards baryonic
$B$ decays suddenly faded away after 1992. This situation was
dramatically changed in the past five years. Interest in this area
was revitalized by many new measurements at CLEO, Belle and BaBar
followed by active theoretical studies.

In this talk we would like to give an overview of the experimental
and theoretical status of exclusive baryonic $B$ decays by the end
of 2005.

\begin{table}[h]
\caption{Experimental upper limits on the branching ratios of
charmless two-body baryonic $B$ decays. } \label{tab:2body}
\begin{ruledtabular}
\begin{tabular}{l l l l  }
 Decay & BaBar \cite{BaBar:pp,BaBar:ppK} & Belle \cite{Belle:2body} & CLEO \cite{CLEO:2body} \\ \hline
 $\ov B^0\to p\bar p$ & $2.7\times 10^{-7}$ & $4.1\times
 10^{-7}$ & $1.4\times 10^{-6}$ \\
 $\ov B^0\to\Lambda\bar\Lambda$ &  & $6.9\times 10^{-7}$ & $1.2\times 10^{-6}$ \\
 $B^-\to\Lambda\bar p$ & & $4.9\times
 10^{-7}$ &  $1.5\times 10^{-6}$  \\
 $B^-\to\Lambda(1520)\bar p$ & $1.5\times 10^{-6}$ & &  \\
 $B^-\to p\bar\Delta^{--}$ & & & $1.5\times 10^{-4}$  \\
 $B^-\to\Delta^0\bar p$ & & & $3.8\times 10^{-4}$  \\
 $\ov B^0\to\Delta^{++}\bar\Delta^{--}$ & & &  $1.1\times 10^{-4}$  \\
 $\ov B^0\to \Delta^0\bar\Delta^0$ & & & $1.5\times 10^{-3}$  \\
\end{tabular}
\end{ruledtabular}
\end{table}

\subsection{Experimental status}
{\bf A.1. Two-body decays}
 \vskip 0.3cm
The experimental results for two-body baryonic $B$ decays  are
summarized in Tables \ref{tab:2body} and \ref{tab:2bodycharm} for
charmless and charmful decays, respectively. It is clear that the
present limit on charmless ones has been pushed to the level of
$10^{-7}$. In contrast, four of the charmful 2-body baryonic $B$
decays have been observed in recent years; among them $\ov
B^0\to\Lambda_c^+\bar p$ is the first observation of the 2-body
baryonic $B$ decay mode \cite{Belle:Lamcp}. The decays with two
charmed baryons in the final state were measured by Belle recently
\cite{2cbaryon}. Taking the theoretical estimates (see e.g. Table
III of \cite{CT93}), $\B(\Xi_c^0\to \Xi^-\pi^+)\approx 1.3\%$ and
$\B(\Xi_c^+\to \Xi^0\pi^+)\approx 3.9\%$ together with the
experimental measurement $\B(\Xi_c^+\to \Xi^0\pi^+)/\B(\Xi_c^+\to
\Xi^-\pi^+\pi^+)=0.55\pm0.16$ \cite{PDG}, it follows that
  \be
 \B(B^-\to\Xi_c^0\bar\Lambda_c^-)\approx 4.8\times 10^{-3}, \qquad
 \B(\ov B^0\to\Xi_c^+\bar\Lambda_c^-)\approx 1.2\times 10^{-3}.
 \en
Therefore, the two-body doubly charmed baryonic $B$ decay $B\to
\B_c\bar \B'_c$ has a branching ratio of order $10^{-3}$. Hence,
we have the pattern
 \be \label{eq:2bodypattern}
 \B_c\bar \B'_c~(\sim 10^{-3})\gg \B_c\bar\B~(\sim 10^{-5})\gg
 \B_1\bar \B_2~(\lsim 10^{-7})
 \en
for two-body baryonic $B$ decays.

\begin{table}[t]
\caption{Branching ratios (in units of $10^{-5}$) of charmful
two-body baryonic $B$ decays. } \label{tab:2bodycharm}
\begin{ruledtabular}
\begin{tabular}{l l l   }
 Decay & Belle \cite{Belle:Lamcp,Belle:Lamcppi,2cbaryon} & CLEO \cite{CLEO:cbaryon} \\ \hline
 $\ov B^0\to\Lambda_c^+\bar p$ &  $2.19^{+0.56}_{-0.49}\pm0.32\pm0.57$  &
 $<9$ \\
 $B^-\to\Lambda_c^+\bar\Delta^{--}$ & $0.65^{+0.56}_{-0.51}\pm0.06\pm0.17~(<1.9)$ & \\
 $B^-\to\Lambda_c^+\bar\Delta_X(1600)^{--}$ & $5.90^{+1.03}_{-0.96}\pm0.55\pm1.53$ & \\
 $B^-\to\Lambda_c^+\bar\Delta_X(2420)^{--}$ & $4.70^{+1.00}_{-0.92}\pm0.43\pm1.22$ & \\
 $\ov B^0\to\Lambda_c(2593)^-/\Lambda_c(2625)^-\bar p$ & $$  & $<11$ \\
 $B^-\to\Sigma_c^0\bar p$ &  $3.67^{+0.74}_{-0.66}\pm0.36\pm0.95$  & $<8$ \\
 $B^-\to\Sigma_c(2520)^0\bar p$ &  $1.26^{+0.56}_{-0.49}\pm0.12\pm0.33~(<2.7)$ & \\
 \hline
 $B^-\to\Xi_c^0(\to\Xi^-\pi^+)\bar\Lambda_c^-$ &
 $4.8^{+1.0}_{-0.9}\pm1.1\pm1.2$ & \\
 $\ov B^0\to\Xi_c^+(\to\Xi^-\pi^+\pi^+)\bar\Lambda_c^-$ &
 $9.3^{+3.7}_{-2.8}\pm1.9\pm2.4$ & \\
\end{tabular}
\end{ruledtabular}
\end{table}

\vspace{0.2cm} {\bf A.2.  Three-body decays}
 \vskip 0.3cm

The measurements of three-body or four-body baryonic $B$ decays
are quite fruitful and many new results have been emerged in the
past years. For the charmless case, Belle \cite{Belle:3charmless}
has observed 6 different modes while BaBar has measured one of
them, see Table \ref{tab:3charmless}. The channel $B^-\to p\bar p
K^-$ announced by Belle nearly four years ago \cite{Belle:ppK} is
the first observation of charmless baryonic $B$ decays. Recently
Belle has studied the baryon angular distribution in the
baryon-antibaryon pair rest frame \cite{Belle:3charmless1}, while
BaBar has measured the Dalitz plot asymmetry in the decay $B^-\to
p\bar pK^-$. These measurements provide valuable information on
the decay dynamics, as we shall discuss later.

\begin{table}[t]
\caption{Branching ratios (in units of $10^{-6}$) of charmless
three-body baryonic $B$ decays. } \label{tab:3charmless}
\begin{ruledtabular}
\begin{tabular}{l l l   }
 Mode~~ & BaBar \cite{BaBar:ppK}  & Belle \cite{Belle:3charmless,Belle:3charmless1}  \\ \hline
 $B^-\to p\bar p K^-$ & $6.7\pm0.5\pm0.4$ &
 $5.30^{+0.45}_{-0.39}\pm0.58$ \\
 $\ov B^0\to p\bar p \ov K^0$ & & $1.20^{+0.32}_{-0.22}\pm0.14$ \\
 $B^-\to p\bar p K^{*-}$  & &  $10.31^{+3.62+1.34}_{-2.77-1.65}$ \\
 $B^-\to p\bar p\pi^-$ &  &  $3.06^{+0.73}_{-0.62}\pm0.37$ \\
 $\ov B^0\to\Lambda\bar p\pi^+$ & & $3.27^{+0.62}_{-0.51}\pm0.39$ \\
 $B^-\to \Lambda\bar\Lambda K^-$ & & $2.91^{+0.90}_{-0.70}\pm0.38$ \\
 $B^-\to \Lambda\bar\Lambda \pi^-$ & & $<2.8$ \\
 $\ov B^0\to p\bar p\ov K^{*0}$ & & $<7.6$ \\
 $\ov B^0\to\Lambda \bar pK^+$ & & $<0.82$ \\
 $\ov B^0\to\Sigma^0\bar p\pi^+$ &  &  $<3.8$ \\
\end{tabular}
\end{ruledtabular}
\end{table}

Table \ref{tab:3charm} summarizes the measured branching ratios of
charmful baryonic decays with one charmed meson or one charmed
baryon or two charmed baryons in the final state.  In general,
Belle results are slightly smaller than the CLEO measurements. The
decay $B^-\to J/\psi\Lambda\bar p$ was first measured by BaBar
\cite{BaBar:JLamp} and an observation of this mode was made by
Belle recently \cite{Belle:JLamp}. We see that $\B(B\to
\B_c\bar\B'_cM)\sim {\cal O}(10^{-3})$ and $\B(B\to \B_c\bar\B
M)\sim {\cal O}(10^{-4})$. The decay $B\to J/\psi\Lambda\bar p$
with the branching ratio of order $10^{-5}$ is suppressed due to
color suppression.

\begin{table}[t]
\caption{Experimental measurements of the branching ratios (in
units of $10^{-4}$) for the $B$ decay modes with a charmed baryon
$\Lambda_c(2285)$ or
$\Lambda_{c1}=\Lambda_c(2593)/\Lambda_c(2625)$ or $\Sigma_c(2455)$
or $\Sigma_{c1}=\Sigma_c(2520)$ or a charmed meson in the final
state.
 } \label{tab:3charm}
\begin{ruledtabular}
\begin{tabular}{l l l l  }
 Mode~~ & BaBar \cite{BaBar:Dpp,BaBar:JLamp} & Belle
 \cite{Belle:3charm,Belle:Lamcppi,Belle:JLamp,Belle:Lamcp2pi,Belle:2LamcK}
 & CLEO  \cite{CLEO:Dpp,CLEO:cbaryon} \\ \hline
 $\ov B^0\to D^{*+}p\bar p\pi^-$ & $5.61\pm0.59\pm0.73$ & &
 $6.5^{+1.3}_{-1.2}\pm 1.0$ \\
 $\ov B^0\to D^+p\bar p\pi^-$ & $3.80\pm0.35\pm0.46$ & &
 \\
 $\ov B^0\to D^{*+}n\bar p$ & & & $14.5^{+3.4}_{-3.0}\pm2.7$ \\
 $\ov B^0\to D^0 p\bar p$ & $1.24\pm0.14\pm0.12$ & $1.18\pm0.15\pm0.16$ & \\
 $\ov B^0\to D^{*0} p\bar p$ & $0.67\pm0.21\pm0.09$ & $1.20^{+0.33}_{-0.29}\pm0.21$ & \\
 $B^-\to D^- p\bar p$ & & $<0.15$ & \\
 $B^-\to D^{*-} p\bar p$ & & $<0.15$ & \\
 \hline
 $B^-\to J/\psi\Lambda\bar p$ & $(11.6^{+7.4+4.2}_{-5.3-1.8})\times
 10^{-2}$ & $(11.6\pm2.8^{+1.8}_{-2.3})\times 10^{-2}$ & \\
 $B^-\to J/\psi \Sigma^0\bar p$ & & $<0.11$ & \\
 $\ov B^0\to J/\psi p\bar p$ & & $<8.3\times 10^{-3}$ & \\
 \hline
 $B^-\to\Lambda_c^+\bar p\pi^-\pi^+\pi^-$ & & &
 $22.5\pm2.5^{+2.4}_{-1.9}\pm5.8$ \\
 $B^-\to\Lambda_c^+\bar p\pi^-\pi^0$ & & $11.0\pm1.2\pm1.9\pm2.9$ &
 $18.1\pm2.9^{+2.2}_{-1.6}\pm4.7$ \\
 $\ov B^0\to\Lambda_c^+\bar p\pi^+\pi^-$ & &
 $10.3\pm0.9\pm1.2\pm 2.7$ &  $16.7\pm1.9^{+1.9}_{-1.6}\pm4.3$  \\
 $B^-\to \Lambda_c^+\bar p\pi^-$ &  &
 $2.01\pm0.15\pm0.20\pm0.52$ & $2.4\pm0.6^{+0.19}_{-0.17}\pm0.6$  \\
 $B^-\to\Lambda_{c1}^{+}\bar p\pi^-$ & & & $<1.9$ \\
 \hline
 $B^-\to \Sigma_c^{++}\bar p\pi^-\pi^-$ & & & $2.8\pm 0.9\pm 0.5\pm
 0.7$  \\
 $B^-\to\Sigma_c^0\bar p\pi^+\pi^-$ & & & $4.4\pm1.2\pm0.5\pm1.1$ \\
 $\ov B^0\to \Sigma_c^{++}\bar p\pi^-$ & &
 $1.15\pm0.22\pm0.14\pm0.30$ & $3.7\pm0.8\pm0.7\pm1.0$  \\
 $\ov B^0\to\Sigma_c^0\bar p\pi^+$ & &
 $0.97\pm0.21\pm0.12\pm0.25$ & $2.2\pm0.6\pm0.4\pm0.6$ \\
 $B^-\to\Sigma_c^0\bar p\pi^0$ & & & $4.2\pm 1.3\pm0.4\pm1.1$ \\
 $\ov B^0\to \Sigma_{c1}^{++}\bar p\pi^-$ & &
 $1.04\pm0.24\pm0.12\pm0.27$ & \\
 $\ov B^0\to\Sigma_{c1}^{0}\bar p\pi^+$ & &
 $0.33\pm0.19\pm0.04\pm0.09$ &  \\
 \hline
 $B^-\to \Lambda_c^+\bar\Lambda_c^-K^-$ & &
 $6.5^{+1.0}_{-0.9}\pm1.1\pm3.4$ & \\
 $\ov B^0\to \Lambda_c^+\bar\Lambda_c^-\ov K^0$ & &
 $7.9^{+2.9}_{-2.3}\pm1.2\pm4.1$ & \\
\end{tabular}
\end{ruledtabular}
\end{table}

\begin{figure}
\centerline{
            {\epsfxsize2.1 in \epsffile{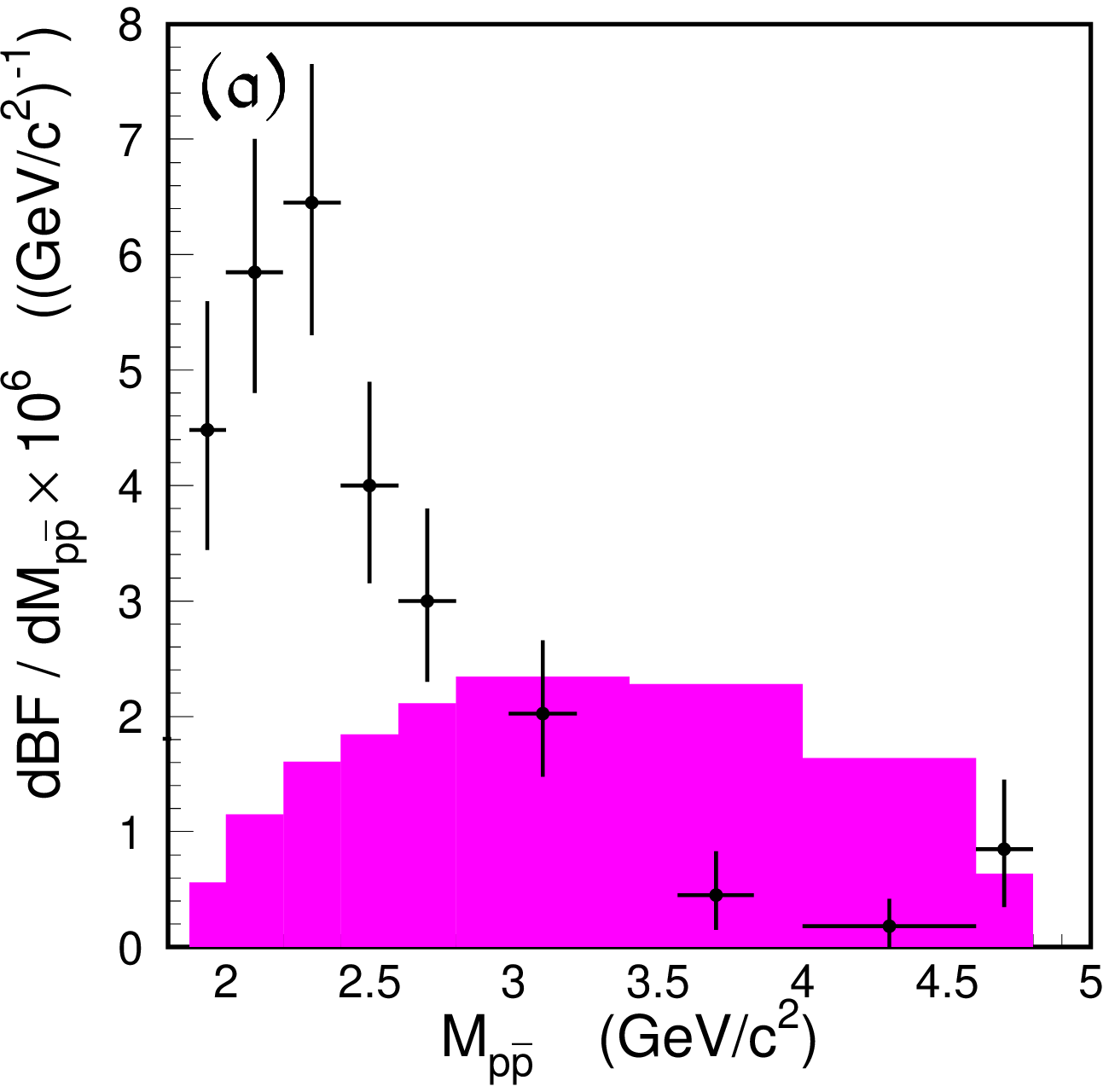}}
            {\epsfxsize2.1 in \epsffile{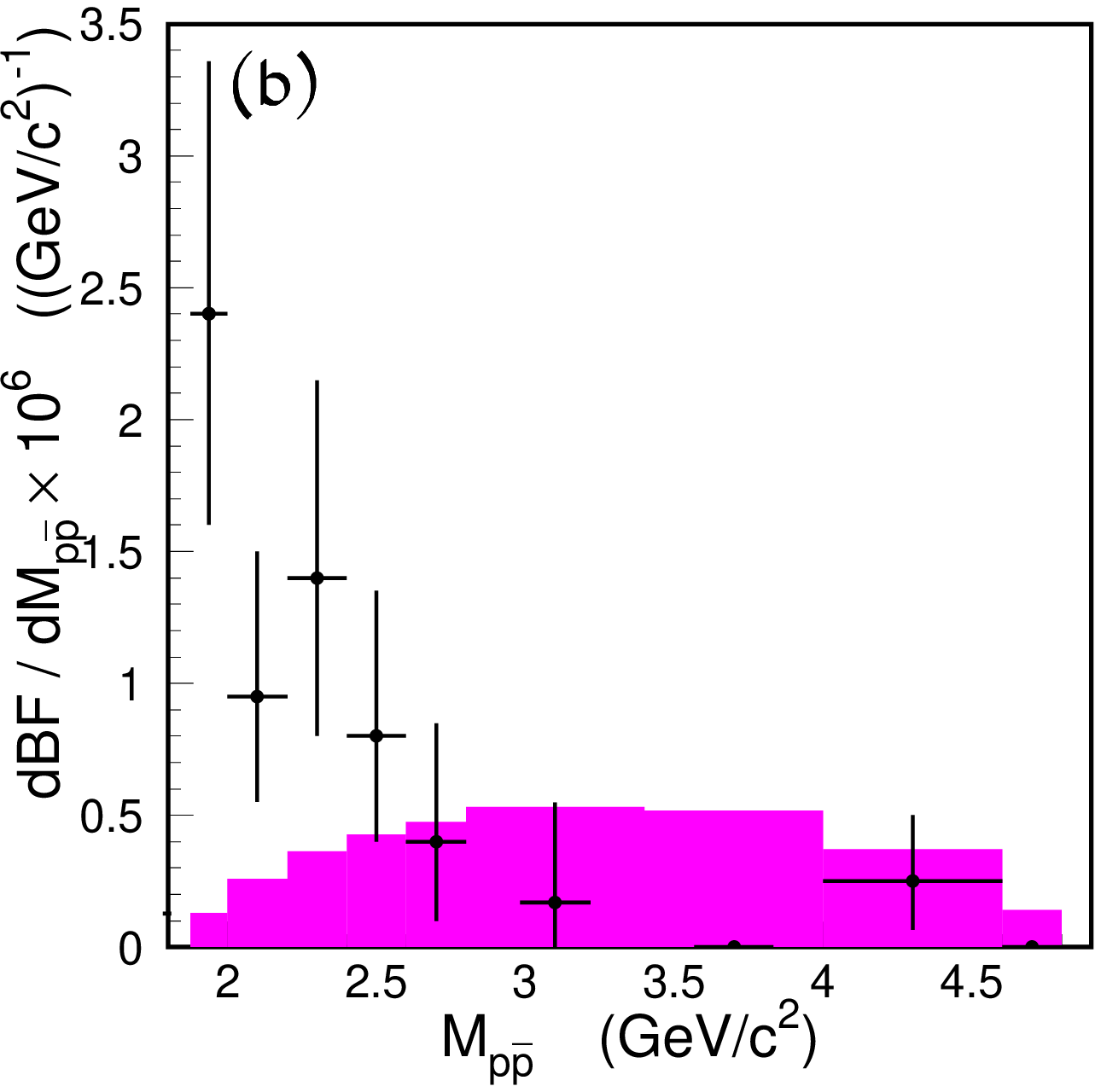}}
            {\epsfxsize2.1 in \epsffile{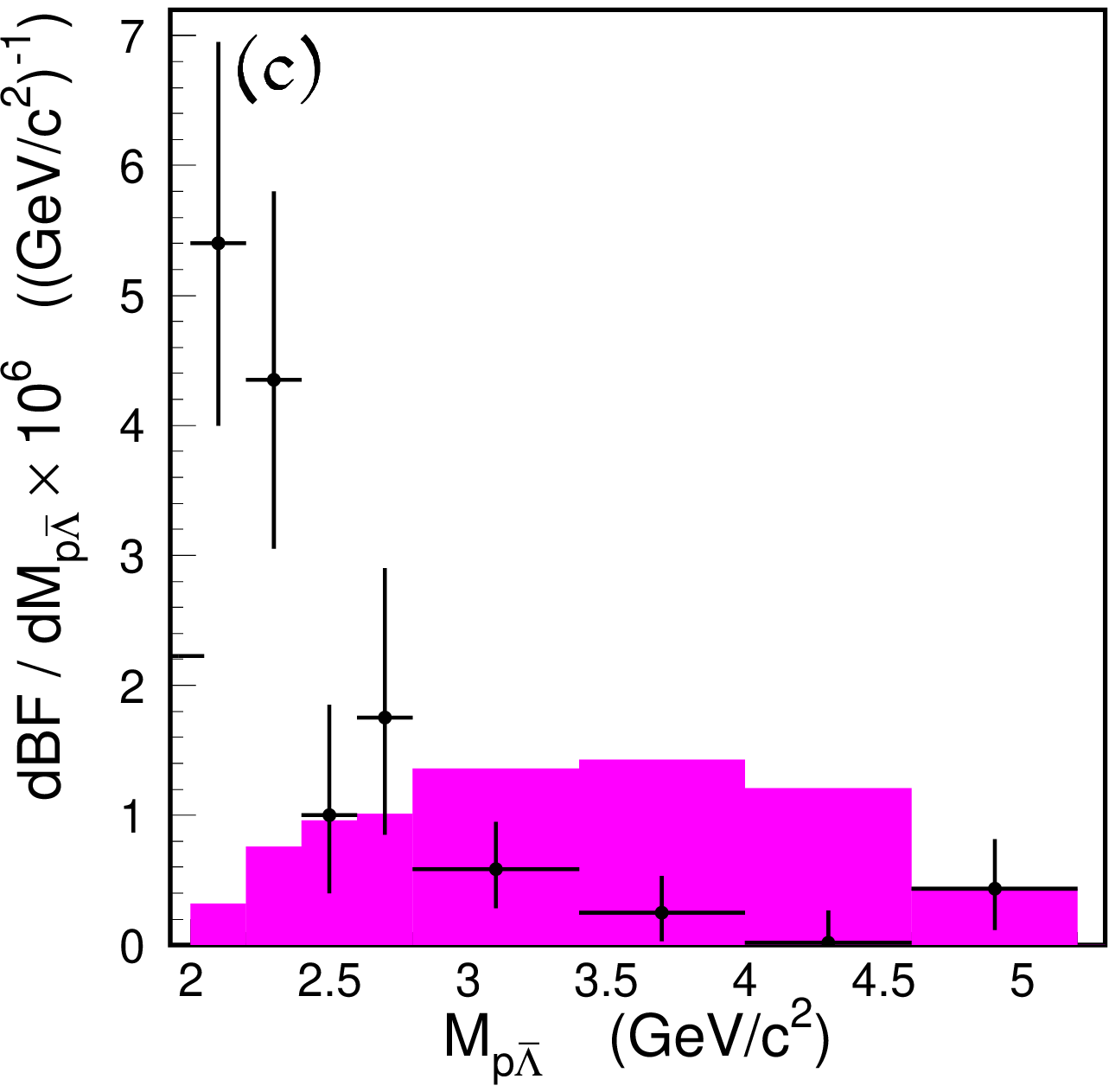}}
            }
\caption{Invariant mass distributions for (a) $p\bar pK^+$, (b)
$p\bar p K_S^0$, and (c) $p\bar \Lambda\pi^-$. The shaded
distribution shows the expectation from a phase-space MC
simulation with area scaled to the signal yield
\cite{Belle:3charmless1}.} \label{fig:spectrum}
\end{figure}

There are two common and unique features for three-body $B\to
\B_1\ov \B_2 M$ decays: (i) The baryon-antibaryon invariant mass
spectrum is peaked near the threshold area (see Fig.
\ref{fig:spectrum}), and (ii) many three-body final states have
rates larger than their two-body counterparts; that is,
$\Gamma(B\to\B_1\ov \B_2M)>\Gamma(B\to\B_1\ov \B_2)$. The low-mass
enhancement effect indicates that the $B$ meson is preferred to
decay into a baryon-antibaryon pair with low invariant mass
accompanied by a fast recoil meson. As for the above-mentioned
second feature, it is by now well established experimentally that
 \be
 \B(B^-\to p\bar p K^-) \gg  \B(\ov B^0\to p\bar p), &\qquad&
 \B(\ov B^0\to \Lambda\bar p\pi^-) \gg  \B(B^-\to \Lambda\bar p),
 \non \\
 \B(B^-\to\Lambda_c^+\bar p\pi^-) \gg \B(\ov B^0\to\Lambda_c^+\bar
 p),  &\qquad&
 \B(B^-\to\Sigma_c^0\bar p\pi^0) \gg \B(B^-\to\Sigma_c^0\bar p).
 \en
This phenomenon can be understood in terms of the threshold
effect, namely, the invariant mass of the dibaryon is preferred to
be close to the threshold. The configuration of the  two-body
decay $B\to\B_1\ov \B_2$ is not favorable since its invariant mass
is $m_B$. In $B\to \B_1\ov\B_2 M$ decays, the effective mass of
the baryon pair is reduced as the emitted meson can carry away
much energies.

An enhancement of the dibaryon invariant mass near threshold has
been observed in the charmless decays $\ov B^0\to\Lambda\bar
p\pi^+$,  $\ov B^0\to p\bar p K_S$, $B^-\to p\bar p K^-$, $B^-\to
p\bar p\pi^-$, $B^-\to\Lambda\bar\Lambda K^-$
\cite{Belle:3charmless,Belle:3charmless1,BaBar:ppK}, and in the
charmful decays $B^-\to\Lambda_c^+\bar p\pi^-$, $\ov B^0\to p\bar
p D^0$ and $\ov B^0\to p\bar p D^{*0}$
\cite{Belle:Lamcppi,Belle:3charm,BaBar:Dpp}. The same threshold
behavior has also been observed in the baryonic $J/\psi$ decays:
$J/\psi\to\gamma p\bar p$ and $J/\psi\to K^-p\bar\Lambda$
\cite{BES03,BES04}. However, no low-mass enhancement effect is
seen in the charmful decays $\ov B^0\to\Sigma_c(2455)^{++}\bar
p\pi^-$, $\Sigma_c(2455)^0\bar p\pi^+$ \cite{Belle:Lamcp2pi} and
$B^-\to J/\psi \Lambda\bar p$ \cite{Belle:JLamp}.

Threshold enhancement was first conjectured by Hou and Soni
\cite{HS}, motivated by the CLEO measurement of $B\to D^*p\bar n$
and $D^*p\bar p\pi$ \cite{CLEO:Dpp}. They argued that in order to
have larger baryonic $B$ decays, one has to reduce the energy
release and  at the same time allow for baryonic ingredients to be
present in the final state. This is indeed the near threshold
effect mentioned above. Of course, one has to understand the
underlying origin of the threshold peaking effect. Hence, the
smallness of the two-body baryonic decay $B\to\B_1\ov\B_2$ has to
do with its large energy release.

Note that the invariant mass distributions for $J/\psi\to\gamma
p\bar p$ and $B^-\to\Lambda_c^+\bar p\pi^-$  are so sharply peaked
near threshold that they can be interpreted as some Breit-Wigner
resonances:\footnote{There is a difference between the threshold
effects observed in $J/\psi\to\gamma p\bar p$ and $B^-\to p\bar
pK^-$ decays. The spectrum for the latter (see Fig.
\ref{fig:spectrum}) is peaked near the threshold, but not really
at threshold, and is much wider than that in $J/\psi$ decays.
Hence, it is often argued that threshold enhancement in
$J/\psi\to\gamma p\bar p$ cannot be explained in terms of the
heavy quark decay process. Nevertheless, it will be interesting to
see if the threshold effects observed in $J/\psi\to\gamma p\bar p$
and $B^-\to p\bar pX$ ($X=K^-,\pi^-$) decays share the same
origin.}
$M=(1859^{+~3+~5}_{-10-25})$ MeV and $\Gamma<30$ MeV for the
former \cite{BES03} and $M=(3.35^{+0.01}_{-0.02}\pm0.02)$ GeV and
$\Gamma=(70^{+40}_{-30}\pm40)$ MeV for the latter
\cite{Belle:Lamcppi}. A popular interpretation of the $p\bar p$
threshold enhancement observed in $J/\psi\to\gamma p\bar p$ is to
postulate the existence of a $p\bar p$ bound state, known as
baryonium. Indeed, a new resonance $X(1835)$ was recently observed
by BES in the $\pi^+\pi^-\eta'$ invariant mass \cite{BES05}. It
has a mass $M=1833.7\pm6.1\pm2.7$ MeV and a width
$\Gamma=67.7\pm20.3\pm7.7$ MeV. As for the low $p\bar p$ mass
enhancement observed in $B$ decays, it has been suggested that a
possible glueball resonance with a mass near 2.3 GeV may
contribute to the threshold enhancement behavior for the $p\bar p
K^-$ mode \cite{CHTglueball}. Although this possibility is ruled
out experimentally \cite{Belle:3charmless1}, a gluonic resonant
state near or below the $p\bar p$ threshold still remains
plausible. For decays such as $\ov B^0\to\Lambda\bar p\pi^+$, $\ov
B^0\to D^0p\bar p$ and $B^-\to p\bar p\pi^-$, the threshold
phenomenon cannot be accounted for by the intermediate gluonic
effects. In this case, low mass enhancement should be explained in
terms of the usual heavy decay process (or the so-called
fragmentation process in \cite{Rosner}).

\subsection{Theoretical progress}
Since baryonic $B$ decays involve two baryons, it is extremely
complicated and much involved. Nevertheless, there are some
theoretical progresses in the past five years.

It is known that two-body baryonic $B$ decays are dominated by
nonfactorizable contributions that are difficult to evaluate. This
nonfactorizable effect can be evaluated in the pole model. Using
the MIT bag model to evaluate the weak matrix elements and the
$^3P_1$ model to estimate the strong coupling constants, it is
found in \cite{CYcharmless} and \cite{CKcharm} that the charmless
and charmful two-body decays can be well described. Chang and Hou
\cite{Chang} have generalized the original version of the diquark
model \cite{Ball} to include penguin effects, though no
quantitative predictions are made. In the meantime, a diagrammatic
approach has been developed for both charmful \cite{Luo} and
charmless \cite{Chua2body} decays. A different approach for
analyzing the helicity structure of the charmless two-body
baryonic decays is performed in \cite{Suzuki} with results similar
to \cite{Chua2body}.

\begin{figure}
\centerline{
            {\epsfxsize2.3 in \epsffile{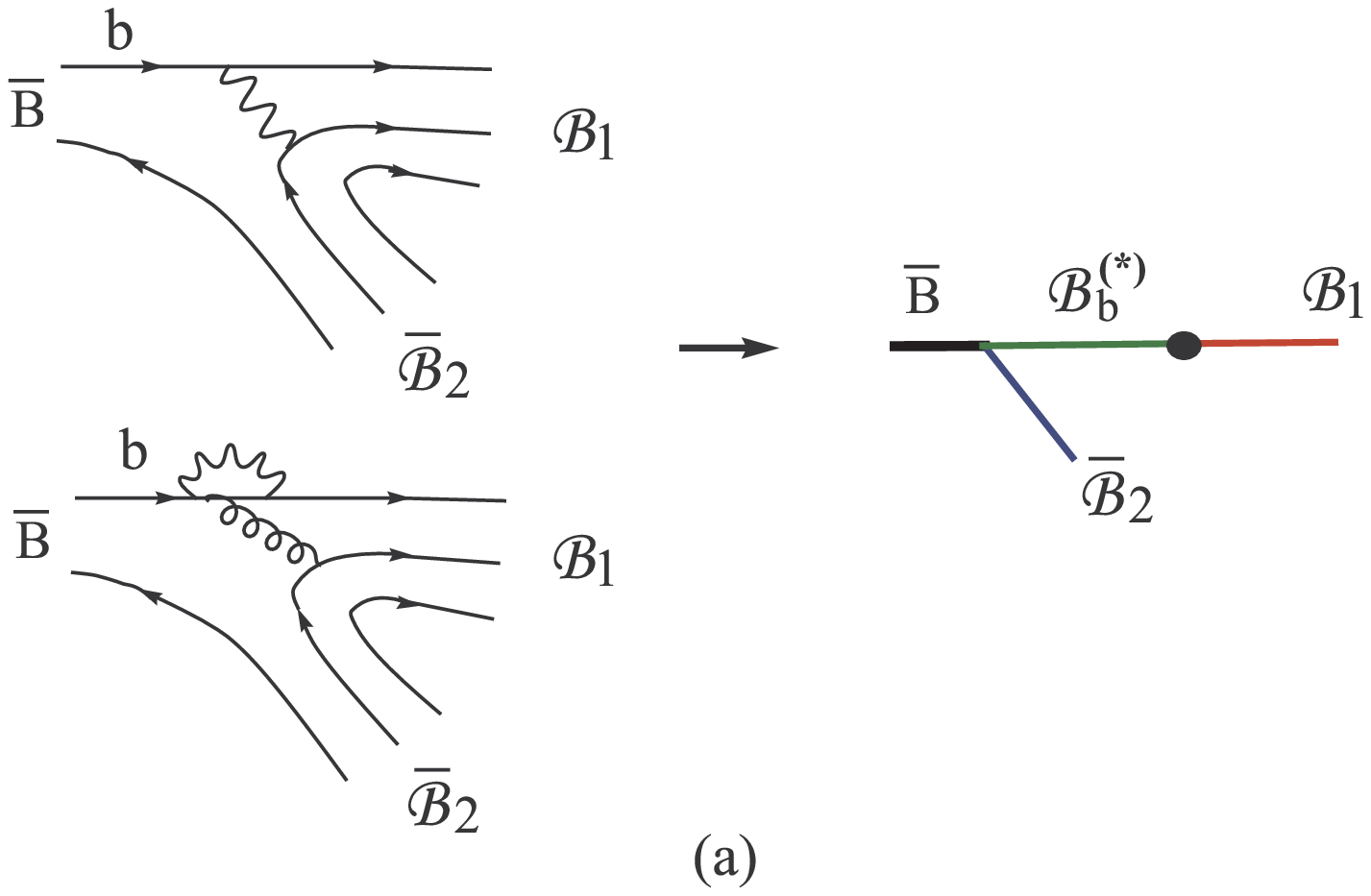}}
            {\epsfxsize2.3 in \epsffile{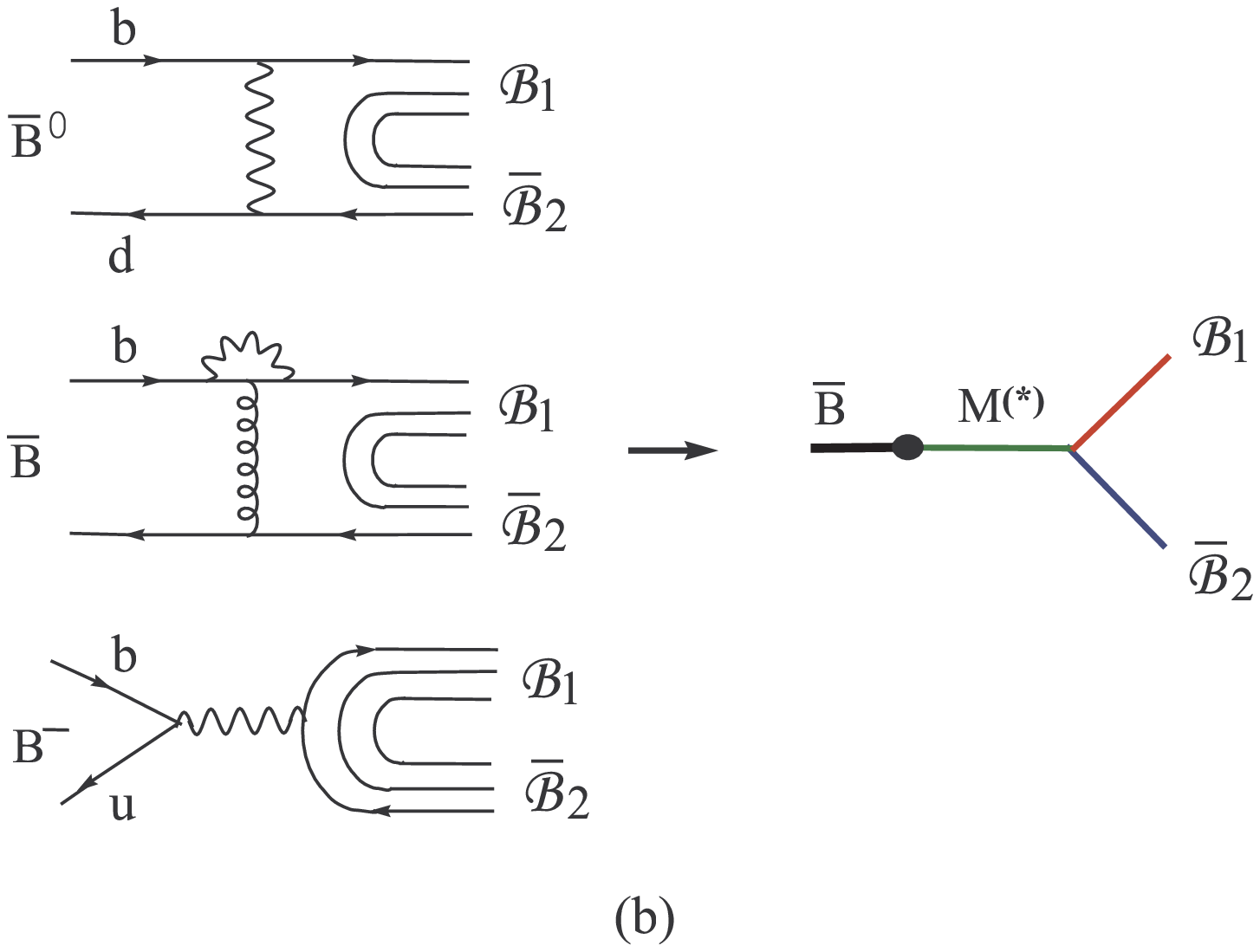}}}
\caption{Quark and pole diagrams for two-body baryonic $B$ decay
    $\ov B\to \B_1\ov \B_2$, where the symbol $\bullet$ denotes the weak
    vertex.} \label{fig:2body}
\end{figure}

Contrary to the two-body baryonic $B$ decay, the three-body decays
do receive factorizable contributions that fall into two
categories: (i) the transition process with a meson emission, $\la
M|(\bar q_3 q_2)|0\ra\la \B_1\ov \B_2|(\bar q_1b)|B\ra$, and (ii)
the current-induced process governed by the factorizable amplitude
$\la \B_1\ov \B_2|(\bar q_1 q_2)|0\ra \la M|(\bar q_3 b)|B\ra$.
The two-body matrix element $\la \B_1\ov \B_2|(\bar q_1 q_2)|0\ra$
in the latter process can be either related to some measurable
quantities or calculated using the quark model. The
current-induced contribution to three-body baryonic $B$ decays has
been discussed in various publications \cite{CHT01,CHT02,CH03}. On
the contrary, it is difficult to evaluate the three-body matrix
element in the transition process and in this case one can appeal
to the pole model \cite{CYcharmless,CKcharm,CKDmeson}.

Weak radiative baryonic $B$ decays $B\to\B_1\ov \B_2\gamma$
mediated by the electromagnetic penguin process $b\to s\gamma$ may
have appreciable rates. Based on the pole model, it is found that
$B^-\to\Lambda\bar p\gamma$ and $B^-\to\Xi^0\bar\Sigma^-\gamma$
have sizable rates and are readily accessible to $B$ factories
\cite{CKrad}.

\section{2-Body baryonic $B$ decays}
As shown in Fig. 2, the quark diagrams for two-body baryonic $B$
decays consist of  internal $W$-emission diagram, $b\to d(s)$
penguin transition, $W$-exchange for the neutral $B$ meson and
$W$-annihilation for the charged $B$. Just as mesonic $B$ decays,
$W$-exchange and $W$-annihilation are expected to be helicity
suppressed. Therefore, the two-body baryonic $B$ decay
$B\to\B_1\ov \B_2$ receives the main contributions from the
internal $W$-emission diagram for tree-dominated modes and the
penguin diagram for penguin-dominated processes. It should be
stressed that, contrary to mesonic $B$ decays, internal $W$
emission in baryonic $B$ decays is not necessarily color
suppressed. This is because the baryon wave function is totally
antisymmetric in color indices. One can see from Fig. 2 that there
is no color suppression for the meson production. In the effective
Hamiltonian approach, the relevant four-quark operators are
$O_1=(\bar sq)(\bar qb)$ and $O_2=(\bar qq)(\bar sb)$. The
combination of the operators $O_1-O_2$ is antisymmetric in color
indices (more precisely, it is a color sextet). Therefore, the
Wilson coefficient for tree-dominated internal $W$-emission would
be $c_1-c_2$ rather than $a_2=c_2+c_1/3$. This is indeed the case
found in the pole model calculation in \cite{CYcharmless}.

From the previous argument that one has to reduce the energy
release in order to have larger baryonic $B$ decays \cite{HS}, it
is expected that
 \be
 \Gamma(B\to\B_1\bar\B_2)=|{\rm CKM}|^2/f({\rm energy~release}),
 \en
where CKM stands for the relevant CKM angles. For charmful modes,
the CKM angles for $\Xi_c\bar\Lambda_c$ and $\Lambda_c\bar p$ have
the same magnitudes except for a sign difference. Therefore, one
will expect
 \be
 \B(\ov B^0\to\Lambda_c^+\bar p)=\B(\ov
 B^0\to\Xi_c^+\bar\Lambda_c^-)({\rm dynamical~suppression}).
 \en
where the dynamical suppression arises from the larger energy
release in $\Lambda_c^+\bar p$ than in $\Xi_c\bar\Lambda_c$. Eq.
(\ref{eq:2bodypattern}) implies that the dynamical suppression
effect is of order $10^{-2}$. Likewise,
 \be
 \B(B^-\to\Lambda\bar p) &=& \B(\ov
 B^0\to\Lambda_c^+\bar p)|V_{ub}/V_{cb}|^2({\rm
 dynamical~suppression})' \non \\
 &\sim& 2\times 10^{-7}({\rm dynamical~suppression})'.
 \en
If the dynamical suppression of $\Lambda\bar p$ relative to
$\Lambda_c\bar p$ is similar to that of $\Lambda_c\bar p$ relative
to $\Xi_c\bar\Lambda_c$, the branching ratio of the charmless
two-body baryonic $B$ decays can be even as small as $10^{-9}$. If
it is the case, then it will be hopeless to see any charmless
two-body baryonic $B$ decays.

Since $B\to\B_1\bar\B_2$ amplitudes are nonfactorizable in nature,
it is very difficult to evaluate them directly.  In order to
circumvent this difficulty, it is customary to assume that the
decay amplitude at the hadron level is dominated by the pole
diagrams with low-lying one-particle intermediate states. The
general amplitude reads
  \be
 {\cal A}(B\to \B_1\ov \B_2)=\bar u_1(A+B\gamma_5)v_2,
 \en
where $A$ and $B$ correspond to $p$-wave parity-violating (PV) and
$s$-wave parity-conserving (PC) amplitudes, respectively. In the
pole model, PC and PV amplitudes are dominated by ${1\over 2}^+$
ground-state intermediate states and ${1\over 2}^-$ low-lying
baryon resonances, respectively.  This pole model has been applied
successfully to nonleptonic decays of hyperons and charmed baryons
\cite{CT92,CT93}. In general, the pole diagram leads to
 \be \label{eq:AandB}
 A=-\sum_{\B_b^*}{g_{\B_b^{*}\to B\B_2}\,b_{\B_b^*\B_1}\over
 m_{1}-m_{\B_b^*} }, \qquad\quad B=\sum_{\B_b}{g_{\B_b\to
 B\B_2}\,
 a_{\B_b\B_1}\over m_{1}-m_{\B_b}}.
 \en

There are two unknown quantities in the above equation: weak
matrix elements and strong couplings. For the former one can
employ the MIT bag model to evaluate the baryon-to-baryon
transitions \cite{CYcharmless}. For the latter, there are two
distinct models for quark pair creation: (i) the $^3P_0$ model in
which the $q\bar q$ pair is created from the vacuum with vacuum
quantum numbers. Presumably it works in the nonperturbative low
energy regime, and (ii) the $^3S_1$ model in which the quark pair
is created perturbatively via one gluon exchange with one-gluon
quantum numbers $^3S_1$. Since the light baryons produced in
two-body baryonic $B$ decays are very energetic, it appears that
the $^3S_1$ model may be more relevant for charmless decays. It is
not difficult to see from Fig. 2(a) that one needs to attach two
hard gluons for charmless and singly charmful 2-body baryonic $B$
decays.

\begin{table}[t]
\caption{Predictions of the branching ratios for some charmless
two-body baryonic $B$ decays classified into two categories:
tree-dominated and penguin-dominated. Branching ratios denoted by
``$\dagger$" are calculated only for the parity-conserving part.
Experimental limits are taken from Table \ref{tab:2body}.}
\label{tab:charmless2body}
\begin{ruledtabular}
\begin{tabular}{l c c c c  } & \cite{Chernyak} & \cite{Jarfi} & \cite{CYcharmless} & Expt. \\
\hline
 $\ov B^0\to p\bar p$ & $1.2\times 10^{-6}$ & $7.0\times
 10^{-6}$  & $1.1\times 10^{-7\dagger}$
 & $<2.7\times  10^{-7}$ \\
 $\ov B^0\to n\bar n$ & $3.5\times 10^{-7}$ & $7.0\times
 10^{-6}$ &  $1.2\times
 10^{-7\dagger}$ & \\
 $B^-\to n\bar p$ &  $6.9\times 10^{-7}$ & $1.7\times
 10^{-5}$ & $5.0\times  10^{-7}$ & \\
 $\ov B^0\to\Lambda\bar\Lambda$ &  & $2\times 10^{-7}$ & $0^\dagger$ &
 $<6.9\times 10^{-7}$ \\
 $B^-\to p\bar \Delta^{--}$ &  $2.9\times 10^{-7}$
 &  $3.2\times 10^{-4}$
 &  $1.4\times 10^{-6}$ & $<1.5\times 10^{-4}$ \\
 $\ov B^0\to p\bar\Delta^-$ &  $7\times 10^{-8}$ & $1.0\times
 10^{-4}$ & $4.3\times 10^{-7}$ & \\
 $B^-\to n\bar\Delta^-$ &
 & $1\times 10^{-7}$ & $4.6\times  10^{-7}$ & \\
 $\ov B^0\to n\bar\Delta^0 $ &  & $1.0\times 10^{-4}$ & $4.3\times 10^{-7}$ & \\
 \hline
 $B^-\to\Lambda\bar p$ &  $\lsim 3\times 10^{-6}$ & & $2.2\times 10^{-7\dagger}$ &
 $<4.9\times 10^{-7}$ \\
 $\ov B^0\to \Lambda\bar n$ & & & $2.1\times 10^{-7\dagger}$ & \\
 $\ov B^0\to\Sigma^+\bar p$ &   $6\times 10^{-6}$ & &  $1.8\times
 10^{-8\dagger}$ & \\
 $B^-\to\Sigma^0\bar p$ &  $3\times 10^{-6}$ & &  $5.8\times 10^{-8\dagger}$ & \\
 $B^-\to\Sigma^+\bar\Delta^{--}$ &  $6\times 10^{-6}$ & &  $2.0\times 10^{-7}$ & \\
 $\ov B^0\to\Sigma^+\bar\Delta^-$ &  $6\times 10^{-6}$ & &  $6.3\times 10^{-8}$ & \\
 $B^-\to\Sigma^-\bar\Delta^0$ &  $2\times 10^{-6}$ & &  $8.7\times 10^{-8}$ & \\
\end{tabular}
\end{ruledtabular}
\end{table}

\begin{table}[t]
\caption{Predictions of charmful two-body baryonic $B$ decays.
Experimental results are taken from Table \ref{tab:2bodycharm}.}
\label{tab:charm2body}
\begin{ruledtabular}
\begin{tabular}{l c c l  } & \cite{Jarfi} & \cite{CKcharm} & Expt.
\\ \hline
 $\ov B^0\to\Lambda_c^+\bar p$ & $1.1\times 10^{-3}$
 & $1.1\times 10^{-5}$ & $(2.19\pm0.84)\times 10^{-5}$ \\
 $B^-\to\Sigma_c^0\bar p$ & $1.5\times 10^{-2}$ & $6.0\times 10^{-5}$ &
 $(3.67^{+0.74}_{-0.66}\pm1.01)\times 10^{-5}$ \\
 $\ov B^0\to\Sigma_c^0\bar n$ & $5.8\times 10^{-3}$ & $6.0\times 10^{-7}$ & \\
 $B^-\to\Lambda_c^+\bar\Delta^{--}$ & $3.6\times 10^{-2}$ & $1.9\times 10^{-5}$ &
 $(0.65^{+0.56}_{-0.51}\pm0.18)\times 10^{-5}(<1.9\times 10^{-5})$ \\
\end{tabular}
\end{ruledtabular}
\end{table}

From Table \ref{tab:charmless2body} we see that the charmless
two-body baryonic decays are predicted at the level of $10^{-7}$.
This may indicate that the dynamical suppression of $\B_1\bar\B_2$
relative to $\B_c\bar\B$ is not significant. The predictions for
charmful $B\to \B_{c}\ov\B$ decays are summarized in Table
\ref{tab:charm2body}. All earlier predictions based on the
sum-rule analysis, the pole model and the diquark model are too
large compared to experiment. Note that we predict that $B^-\to
\Sigma_c^0\bar p$ has a larger rate than $\ov B^0\to \Lambda_c\bar
p$ since the former proceeds via the $\Lambda_b$ pole while the
latter via the $\Sigma_b$ pole and the $\Lambda_b N\bar B$
coupling is larger than the $\Sigma_bN\bar B$ one \cite{CKcharm}.

Since the doubly charmed baryonic decay mode $\Xi_c\bar\Lambda_c$
proceeds via $b\to cs\bar c$, while $\Lambda_c\bar p$ via a $b\to
cd\bar u$ quark transition, the CKM mixing angles for them are the
same in magnitude but opposite in sign, one may wonder why the
$\B_c\bar \B'_c$ mode has a rate two orders of magnitude larger
than $\B_c\bar\B$. Indeed, earlier calculations based on QCD sum
rules \cite{Chernyak} or the diquark model \cite{Ball} all predict
that $\B(B\to \Xi_c\bar\Lambda_c)\approx \B(\ov B\to\B_c\ov N)$,
which is in strong disagreement with experiment. This implies that
some important dynamical suppression effect for the $\B_c\ov N$
production with respect to $\Xi_c\bar\Lambda_c$ is missing in
previous studies. Recently, this issue was investigated in
\cite{CCT}. Since the energy release is relatively small in
charmful baryonic $B$ decay, the $^3P_0$ model for $q\bar q$
production is more relevant. In the work of \cite{CCT}, the
possibility that the $q\bar q$ pair produced via light meson
exchanges such as $\sigma$ and pions is considered. The $q\bar q$
pair created from soft nonperturbative interactions tends to be
soft. For an energetic proton produced in 2-body $B$ decays, the
momentum fraction carried by its quark is large, $\sim {\cal
O}(1)$, while for an energetic charmed baryon, its momentum is
carried mostly by the charmed quark. As a consequence, the doubly
charmed baryon state such as $\Xi_c\bar\Lambda_c$ has a
configuration more favorable than $\Lambda_c\bar p$.

\begin{table}[t]
\caption{Predictions on the branching ratios of
$B^-\to\Xi_c^0\bar\Lambda_c^-$ and $\ov
B^0\to\Xi_c^+\bar\Lambda_c^-$ decays \cite{CCT}. The first and
second errors come from the theoretical uncertainties in the
parameters $\beta$ and $\omega_b$, respectively, which are taken
to be $\beta=1.20\pm0.05$~GeV and $\omega_b=0.45\pm0.05$~GeV.
Results shown in second and third rows are from $\pi$ or $\sigma$
exchange alone, respectively.}
 \label{tab:result}
\begin{ruledtabular}
\begin{tabular}{l r c |l  r c}
  Mode
  & Theory $(10^{-3})$
  & Expt $(10^{-3})$
  & Mode
  & Theory $(10^{-3})$
  & Expt $(10^{-3})$
  \\
\hline
  $\B(B^-\to\Xi_c^0\bar\Lambda_c^-)$
 & $1.0^{+0.3+1.1}_{-0.3-0.7}$
 & $\approx 4.8$
 & $\B(\ov B^0\to\Xi_c^+\bar\Lambda_c^-)$
 & $0.9^{+0.2+1.0}_{-0.3-0.6}$
 & $\approx 1.2$
 \\
 $ \B(B^-\to\Xi_c^0\bar\Lambda_c^-)_{\pi}$
 & $0.8^{+0.2+0.9}_{-0.2-0.6}$
 &
 & $\B(\ov B^0\to\Xi_c^+\bar\Lambda_c^-)_{\pi}$
 & $0.8^{+0.2+0.8}_{-0.2-0.5}$
 &
 \\
 $\B(B^-\to\Xi_c^0\bar\Lambda_c^-)_{\sigma}$
 &$0.1^{+0.0+0.1}_{-0.0-0.1}$
 &
 &$\B(\ov B^0\to\Xi_c^+\bar\Lambda_c^-)_{\sigma}$
 & $0.1^{+0.0+0.1}_{-0.0-0.1}$
 &
 \\
\end{tabular}
\end{ruledtabular}
\end{table}

Assuming that a soft $q\bar q$ quark pair is produced through the
$\sigma$ and $\pi$ meson exchanges in the configuration for $\ov
B\to \Xi_c\bar\Lambda_c$, it is found that its branching ratio is
of order $10^{-3}$ (see \cite{CCT} for detail), in agreement with
experiment. Note that this calculation is not applicable to the
two-body decay $\ov B^0\to\Lambda_c^+\bar p$ with one charmed
baryon in the final state. This is because two hard gluons are
needed to produce an energetic antiproton as noticed before: one
hard gluon for kicking the spectator quark of the $B$ meson to
make it energetic and the other for producing the hard $q\bar q$
pair. The pQCD calculation for this decay will be much more
involved (see e.g. \cite{Li} for pQCD calculations of
$\Lambda_b\to\Lambda J/\psi$). Nevertheless, it is expected that
$\Gamma(\ov B\to \B_c\bar N)\ll \Gamma(\ov
B\to\Xi_c\bar\Lambda_c)$ as the former is suppressed by order of
$\alpha_s^4$. This dynamical suppression effect for the
$\Lambda_c\bar p$ production relative to $\Xi_c\bar\Lambda_c$ has
been neglected in the previous studies based on QCD sum rules
\cite{Chernyak} and on the diquark model \cite{Ball}.

\section{3-Body baryonic $B$ decays}

In three-body baryonic $B$ decays, the emission of the meson $M$
can carry away energies in such a way that the invariant mass of
$\B_1\ov\B_2$ becomes smaller and hence it is relatively easier to
fragment into the baryon-antibaryon pair. One can also understand
this feature more concretely by studying the Dalitz plot. Due to
the $V-A$ nature of the $b\to ud\bar u$ process, the invariant
mass of the diquark $ud$ peaks at the highest possible values in a
Dalitz plot for $b\to ud\bar d$ transition \cite{Buchalla}. If the
$ud$ forms a nucleon, then the very massive $udq$ objects will
intend to form a highly excited baryon state such as $\Delta$ and
$N^*$ and will be seen as $N n\pi(n\geq 1)$ \cite{Dunietz}. This
explains the non-observation of the $N\ov N$ final states and why
the three-body mode $N\ov N \pi(\rho)$ is favored. Of course, this
argument is applicable only to the tree-dominated processes.

\begin{figure}[t]
\centerline{
            {\epsfxsize3.0 in \epsffile{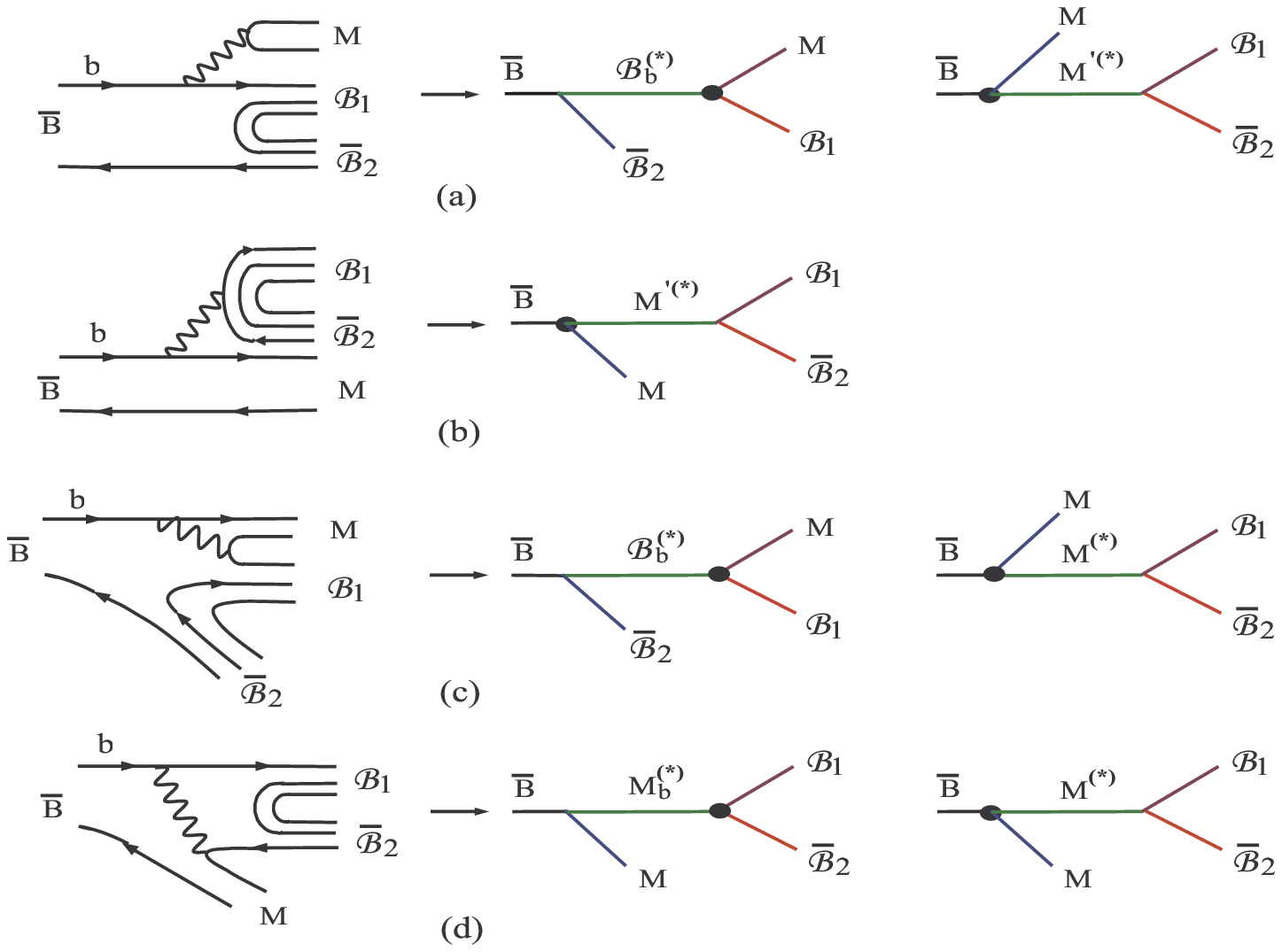}}
            {\epsfxsize2.3 in \epsffile{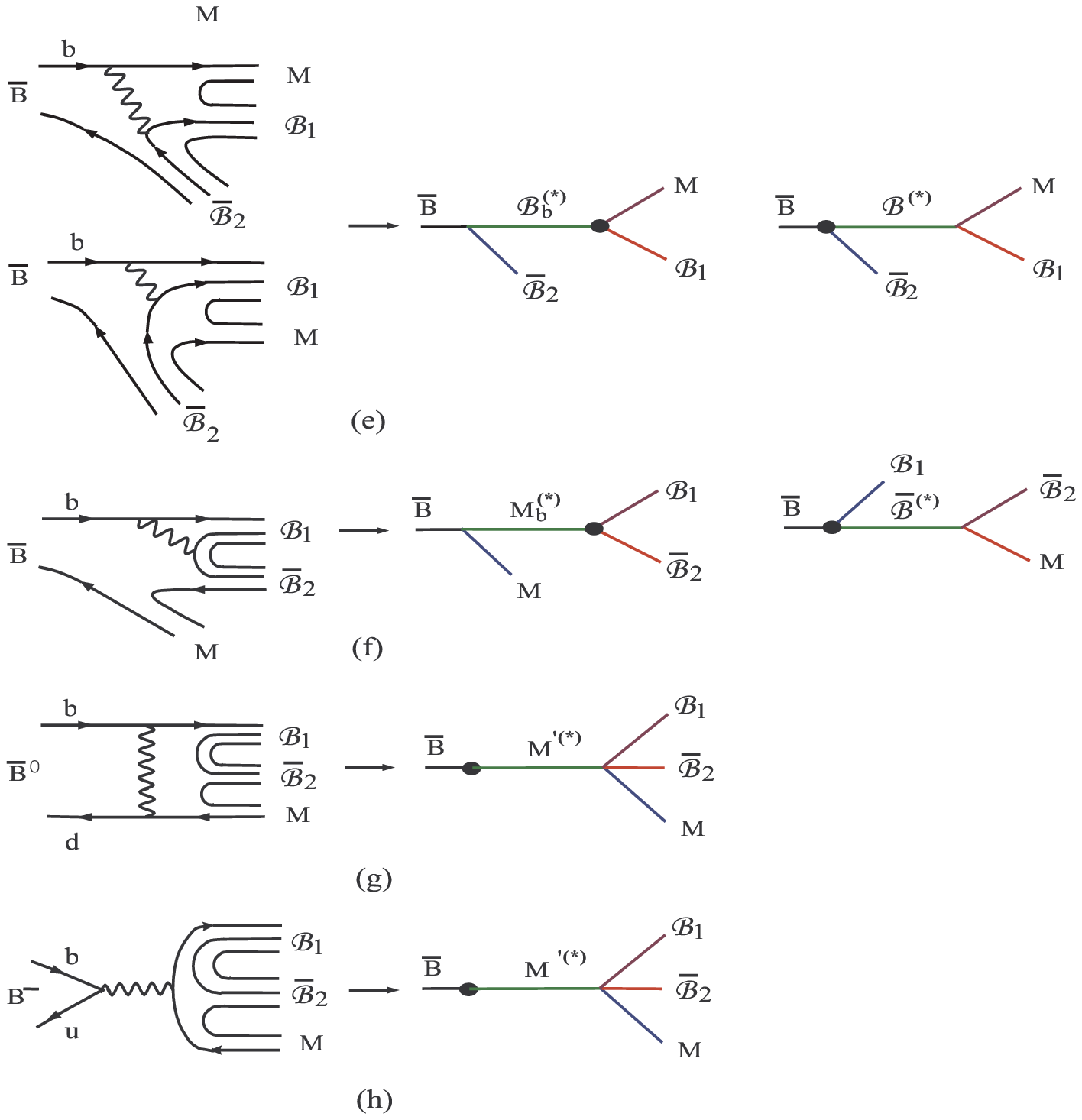}}}
    \caption{ Quark and pole diagrams for three-body baryonic $B$ decay
    $\ov B\to \B_1\ov \B_2M$, where the symbol $\bullet$ denotes the weak
    vertex.} \label{fig:Qdiagrams}
\end{figure}

The quark diagrams and the corresponding pole diagrams for decays
of $B$ mesons to the baryonic final state $\B_1\ov\B_2 M$ are more
complicated. In general, there are two external $W$-diagrams Figs.
3(a)-3(b), four internal $W$-emissions Figs. 3(c)-3(f), and one
$W$-exchange Fig. 3(g) for the neutral $B$ meson and one
$W$-annihilation Fig. 3(h) for the charged $B$. For simplicity,
penguin diagrams are not drawn in Fig. 3, but they can be obtained
from Figs. 3(c)-3(g) by replacing the $b\to u$ tree transition  by
the $b\to s(d)$ penguin transition. Under the factorization
hypothesis, the relevant factorizable amplitudes are
 \be \label{eq:fact}
 {\rm Figs.~3(a),3(c)}: && A\propto \la M|(\bar q_3 q_2)|0\ra\la \B_1\ov
 \B_2|(\bar q_1b)|\ov B\ra,  \non \\
 {\rm Figs.~3(b),3(d)}:&& A\propto \la \B_1\ov \B_2|(\bar q_1 q_2)|0\ra
 \la M|(\bar q_3 b)|\ov B\ra,  \\
 {\rm Figs.~3(g),3(h)}: && A\propto \la\B_1\ov \B_2 M|(\bar
 q_1q_2)|0\ra\la 0|(\bar q_3 b)|\ov B\ra.  \non
 \en

Neglecting the factorizable annihilation contributions which are
helicity suppressed, the three-body decays that receive
factorizable contributions fall into two categories: (i) the
transition process with a meson emission, $\la M|(\bar q_3
q_2)|0\ra\la \B_1\ov \B_2|(\bar q_1b)|\ov B\ra$, and (ii) the
current-induced process governed by the factorizable amplitude
$\la \B_1\ov \B_2|(\bar q_1 q_2)|0\ra \la M|(\bar q_3 b)|\ov
B\ra$. The two-body matrix element $\la \B_1\ov \B_2|(\bar q_1
q_2)|0\ra$ in the latter process can be either related to some
measurable quantities or calculated using the quark model. The
current-induced contribution to three-body baryonic $B$ decays has
been discussed in various publications \cite{CHT01,CHT02,CH03}. On
the contrary, it is difficult to evaluate the three-body matrix
element in the transition process and in this case one can appeal
to the pole model \cite{CYcharmless,CKcharm,CKDmeson}. For Figs.
3(a) and 3(c) we will consider the pole diagrams to evaluate
3-body matrix elements. The 3-body matrix element $\la \B_1\ov
\B_2|(\bar q_1b)|B\ra$ receives contributions from point-like
contact interaction (i.e. direct weak transition) and pole
diagrams \cite{CYcharmless}.

It should be stressed that among the four internal $W$-emission
diagrams, Figs. 3(c) and 3(d) are color suppressed while Figs.
3(e) and 3(f) are not due to baryon wave function antisymmetric in
color indices \cite{CKcharm}. For example, $B\to J/\psi\Lambda\bar
p$ proceeds via Fig. 3(c), while $B^-\to\Sigma_c^0\bar p\pi^+$
receives the dominant contributions from Figs. 3(e) and 3(f). The
experimental observation that $\Sigma_c^0\bar p\pi^+$ has a rate
similar to $\Sigma_c^{++}\bar p\pi^-$ which proceeds via external
$W$-emission Fig. 3(a) and that $J/\psi\Lambda\bar p$ is
suppressed by one order of magnitude implies the color suppression
for Fig. 3(c) and non-suppression for Figs. 3(e) and 3(f).

\subsection{Current-induced three-body baryonic $B$ decays}
Current-induced three-body baryonic $B$ decays such as $\ov
B^0\to\Lambda\bar p\pi^+$ provide an ideal place for understanding
the threshold enhancement effects. Theoretically, the low-mass
enhancement effect is closely linked to the behavior of the baryon
form factors occurred in the vacuum to $\B_1\ov\B_2$ transition
matrix element
 \be
\la \B_1(p_1)\ov\B_2(p_2)|(V\pm A)_\mu|0\ra &=& \bar
u_1(p_1)\Bigg\{f_1^{\B_1\B_2}(t)\gamma_\mu+i{f_2^{\B_1\B_2}(t)\over
m_1+m_2} \sigma_{\mu\nu}q^\nu+{f_3^{\B_1\B_2}(t)\over
m_1+m_2}q_\mu \non  \\ && \pm
\Big[g_1^{\B_1\B_2}(t)\gamma_\mu+i{g_2^{\B_1\B_2}(t)\over m_1+m_2}
\sigma_{\mu\nu}q^\nu+{g_3^{\B_1\B_2}(t)\over
m_1+m_2}q_\mu\Big]\gamma_5\Bigg\}v_2(p_2),
 \en
where $t=(p_1+p_2)^2=M_{\B_1 \bar \B_2}^2$. The form factors
$f_i^{\B_1\B_2}(t)$ and $g_i^{\B_1\B_2}(t)$ expressed in terms of
a power series of the inverse of the dibaryon invariant mass
squared $M_{\B_1\bar\B_2}^2$ will fall off sharply with $t$. For
octet baryons one can apply SU(3) symmetry to relate the vector
form factors $f_i^{\B_1\B_2}$ to the nucleon magnetic and electric
form factors which have been measured over a large range of $q^2$.

The decay $\ov B^0\to\Lambda\bar p\pi^+$ receives the dominant
factorizable contributions from the tree diagram Fig. 3(b) and the
penguin diagram Fig. 3(d) with the amplitudes (see e.g.
\cite{CYcharmless})
 \be
 A(\ov B^0\to\Lambda\bar p\pi^+) &=& {G_F\over\sqrt{2}}
\la\pi^+|(\bar ub)|\ov B^0\ra\Big\{
(V_{ub}V^*_{us}a_1-V_{tb}V^*_{ts}a_4)\la\Lambda\bar p|(\bar
su)|0\ra \non \\
&+& 2a_6V_{tb}V^*_{ts}{(p_\Lambda+p_{\bar p})\over
m_b-m_u}\la\Lambda\bar p|\bar s(1+\gamma_5)u|0\ra\Big\}.
 \en
For the matrix element of scalar and pseudoscalar densities,
 \be
 \la \Lambda\bar p|\bar s(1+\gamma_5)u|0\ra  &=&
 f_S(t)\bar u_\Lambda v_{\bar p} +g_P(t)\bar u_\Lambda\gamma_5
 v_{\bar p},
 \en
the form factors $f_S(t)$ and $g_P(t)$ can be related to
$f_1^{\Lambda p}(t)$ and $g_{1,3}^{\Lambda p}(t)$ via equations of
motion. Based on the pQCD counting rule \cite{Brodsky} which gives
rise to the leading power in the large-$t$ fall-off of the form
factor by counting the number of gluon exchanges necessary to
distribute the large momentum transfer to all constituents, the
form factor generally has the asymptotic form
 \be
 F(t)\to {a\over t^2}+{b\over t^3}
 \en
in the limit of large $t$, where
$F(t)=f_i(t),g_i(t),f_S(t),g_P(t)$.

\begin{figure}[t]
\vspace{0cm} \centerline{
            {\epsfxsize1.9in \epsffile{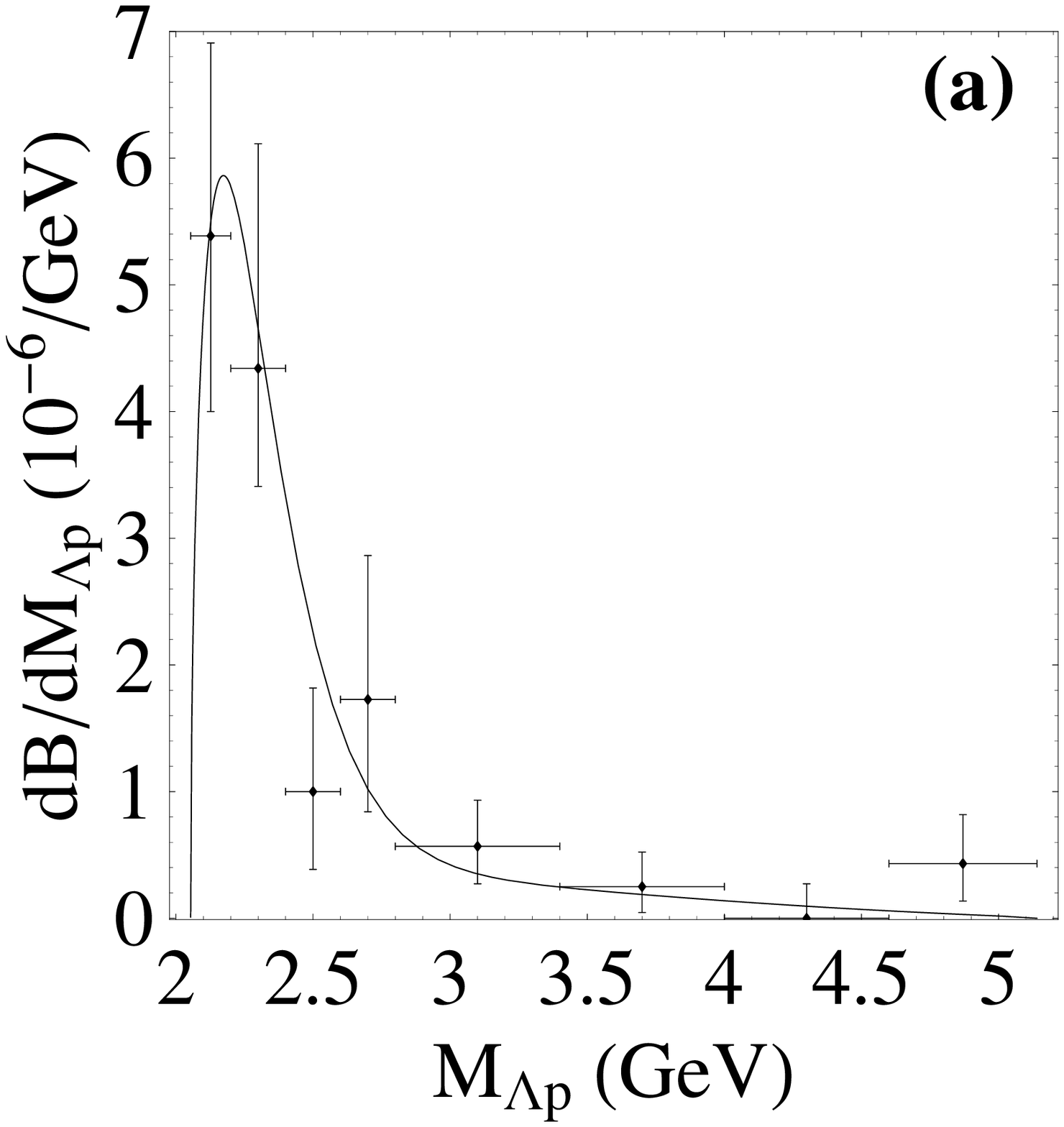}}
            {\epsfxsize2.1in \epsffile{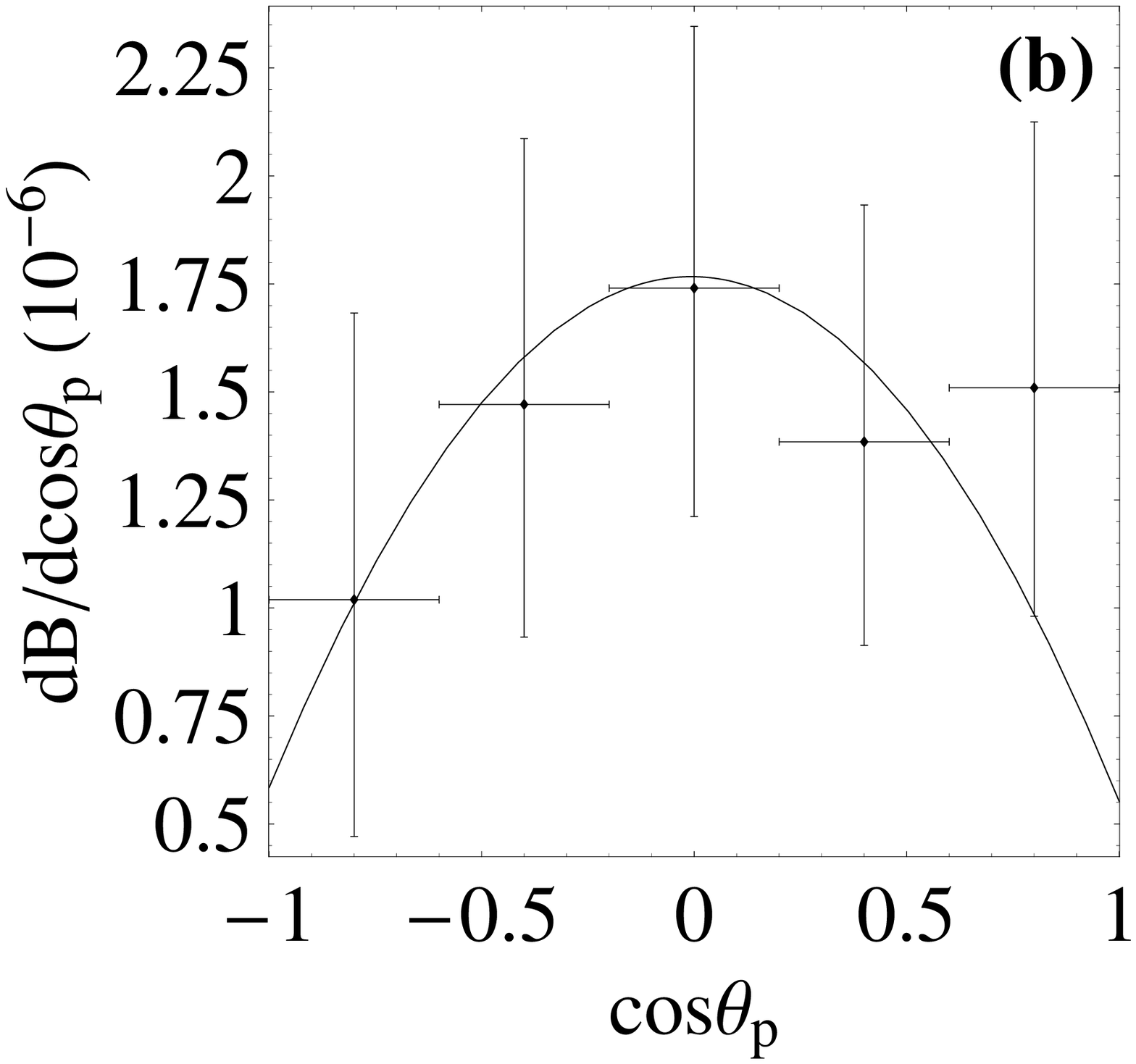}}}
 \vskip0.2cm
\caption{(a) $\Lambda\bar p$ invariant mass spectrum and (b) the
antiproton angular distribution in the $\Lambda\bar p$ rest frame
for the decay $\ov B^0\to\Lambda\bar p\pi^+$, where $\theta_p$ is
the angle between the antiproton direction and the pion direction
in the $\Lambda \bar p$ rest frame. Experimental data are taken
from \cite{Belle:3charmless1} and the theoretical curves are from
\cite{Tsai}.} \label{fig:Lamppi}
\end{figure}

Fig. \ref{fig:Lamppi} shows the dibaryon mass spectrum and the
angular distribution in the $\Lambda\bar p$ rest frame. The
threshold enhancement effect depicted in Fig. \ref{fig:Lamppi}(a)
is closely related to the asymptotic behavior of various form
factors, namely, they fall off fast with the dibaryon invariant
mass. A detailed study in \cite{Tsai} shows that the differential
decay rate for $\Lambda\bar p\pi^+$ should be in the form of a
parabola that opens downward. This is indeed confirmed by
experiment (Fig. \ref{fig:Lamppi}(b)) where it is clear that the
pion has no preference for its correlation with the $\Lambda$ or
the $\bar p$. This feature can be understood as follows. Since the
dibaryon invariant mass in the $B$ rest frame is given by
 \be
 M_{\Lambda \bar p}^2=m_\Lambda^2+m_p^2+2(E_\Lambda E_{\bar
 p}-|\vec{p}_\Lambda||\vec{p}_{\bar p}|\cos\theta_{\Lambda\bar
 p}),
 \en
threshold enhancement implies that the baryon pair $\Lambda$ and
$\bar p$ tend to move collinearly in this frame, i.e.
$\theta_{\Lambda\bar p}\to 0$. In the penguin diagram responsible
for $\ov B^0\to\Lambda\bar p\pi^+$, both $\Lambda$ and $\bar p$
pick up energetic $s$ and $\bar u$ quarks, respectively, from the
$b$ decay.  When the system is boosted to the $\Lambda\bar p$ rest
frame, the pion is moving away from the dibaryon system.
Therefore, the distribution should be symmetric. In contrast, the
argument in \cite{Rosner} that the $\bar p$ and $\pi^+$ are
neighbors in the fragmentation chain so that the $\pi^+$ is
correlated more strongly to the $\bar p$ than to the $\Lambda$ (or
$\la m_{\bar p\pi}\ra<\la m_{\Lambda\pi}\ra$)  will lead to an
asymmetric angular distribution. Evidently, this feature is not
borne out by experiment.

\subsection{Decays involving $B\to\B_1\ov \B_2$ transition} Apart
from the purely transition-induced decays such as $\ov B^0\to
D^{(*)0}p\bar p$, $\ov B^0\to\Sigma_c^{++}\bar p\pi^-$, most
decays receive both current- and transition-induced contributions,
e.g. $B^-\to p\bar p K^-,p\bar p\pi^-,\Lambda\bar\Lambda
K^-,\Lambda_c^+\bar p\pi^-$ and $\ov B^0\to p\bar p K_S$. The
phenomenon of threshold enhancement has been observed in all these
modes.

The factorizable amplitude for the transition-induced process $\la
M|(\bar q_3 q_2)|0\ra\la \B_1\ov \B_2|(\bar q_1b)|\ov B\ra$ can be
further simplified to
 \be
 \la P|(V\pm A)_\mu|0\ra\la \B_1\ov\B_2|(V-A)^\mu|\ov B\ra=
 \pm if_P\la \B_1\ov \B_2|S+ P|\ov B\ra
 \en
for the case that the emitted meson is a pseudoscalar one. In this
case, the matrix element of $\ov B\to\B_1\bar\B_2$ can be
parametrized in terms of four unknown 3-body form factors as
 \be \label{eq:3bodyme}
 \la \B_1\ov\B_2|S\pm P|\ov B\ra=i\bar
 u\left[F_A^{\ov B\to\B_1\ov\B_2} p\!\!\!/\gamma_5+F_
 P^{\ov B\to\B_1\ov\B_2}\gamma_5\pm(F_V^{\ov B\to\B_1\ov\B_2}
 p\!\!\!/+F_S^{\ov B\to\B_1\ov\B_2})\right]v.
 \en

\begin{figure}[t]
\centerline{ {\epsfxsize1.7 in \epsffile{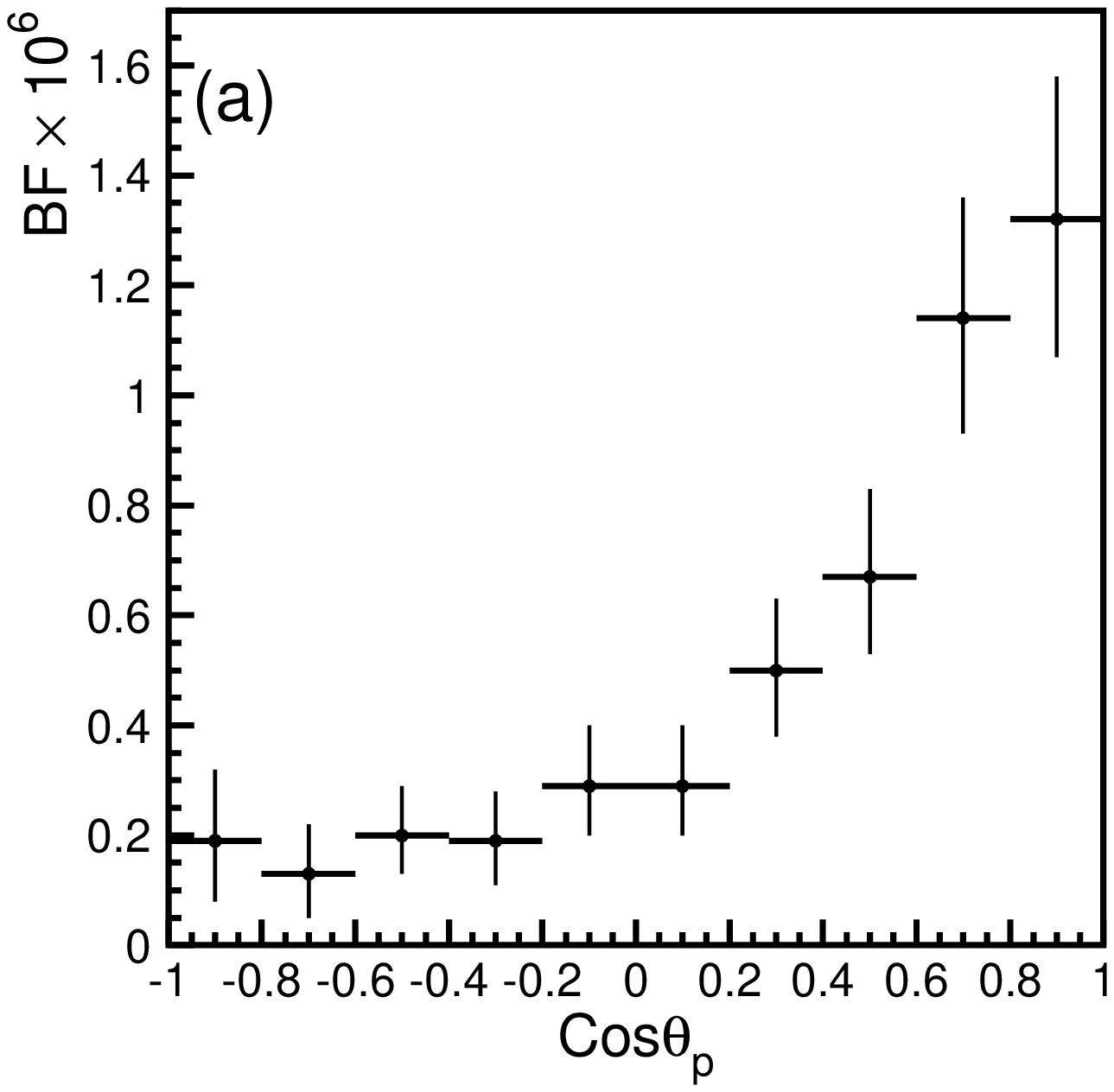}}
            {\epsfxsize2.0 in \epsffile{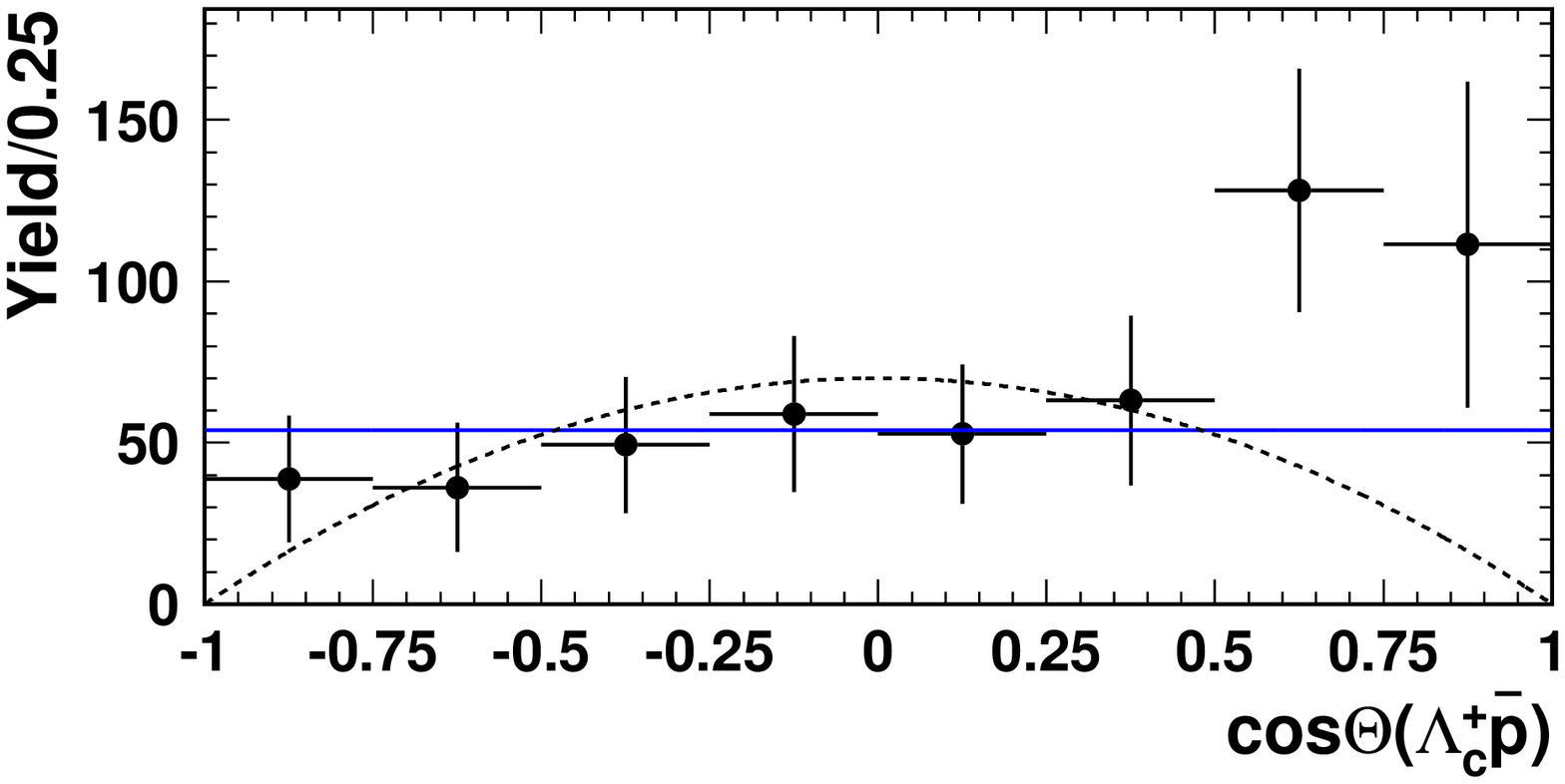}}
            {\epsfxsize1.7 in \epsffile{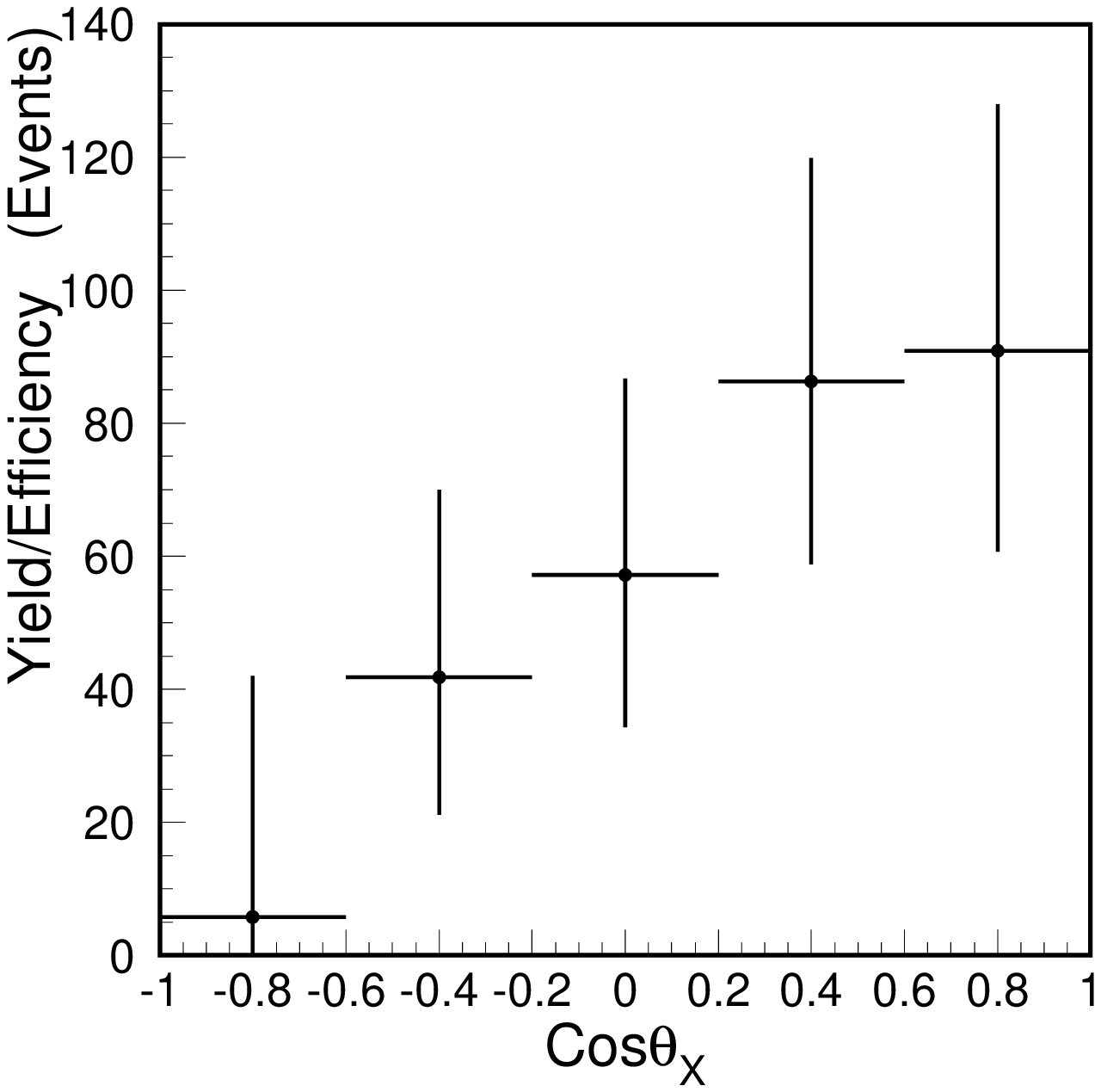}}}
 \centerline{\,\,(a)\hspace{4.0cm}(b)\hspace{4.0cm}(c)}
            \caption{Angular distributions of (a) the proton for $B^-\to p\bar pK^-$
\cite{Belle:3charmless} and the antiproton for (b)
$\Lambda_c^+\bar p\pi^-$ \cite{Belle:Lamcppi} and (c) $\Lambda\bar
p\gamma$ \cite{Belle:Lampgam} modes. Measurements are done in the
dibaryon rest frame. $\theta_p$ in (a) is the angle between the
proton direction and the kaon direction, while $\Theta$ in (b)
($\theta_X$ in (c)) is the angle between the antiproton and the
pion (photon). } \label{fig:angular}
\end{figure}

It has been argued in \cite{CHT02} that in the asymptotic
$t\to\infty$ limit, the pQCD counting rule implies
 \be \label{eq:3bodyFF}
 F_{V,A}^{\ov B\to\B_1\ov\B_2}\to {1\over t^3},\qquad\qquad
 F_P^{\ov B\to\B_1\ov\B_2}\to {1\over t^4}
 \en
as it needs three hard gluon exchanges to distribute the large
momentum transfer released from the $b\to q$ transition.
Consequently, just as the current-induced processes, the threshold
enhancement effect is linked to the asymptotic behavior of the
form factors. However, there are two reasons that the momentum
dependence of the 3-body form factors cannot depend solely on the
dibaryon invariant mass as proposed by Eq. (\ref{eq:3bodyFF}).
First, no threshold enhancement is observed in the decay mode
$\Sigma_c(2455)^{++}\bar p\pi^-$ \cite{Belle:Lamcp2pi}. Second,
the momentum dependence of form factors on the dibaryon mass will
lead to a symmetric angular distribution similar to Fig.
\ref{fig:Lamppi}(b), which is in sharp contradiction to experiment
for e.g. $B^-\to p\bar p K^-$, $B^-\to\Lambda_c^+\bar p\pi^-$ and
$\ov B^0\to\Lambda\bar p\gamma$ where the angular distributions
are obviously asymmetric (Fig. \ref{fig:angular}). In short, for
the 3-body matrix element $\la \B_1(p_1)\ov\B_2(p_2)|S\pm P|\ov
B(p_B)\ra$, the form factors in general depend not only on the
invariant mass $t=(p_1+p_2)^2$ of the dibaryon but also on the
momentum transfer $(p_1+p_3)^2$ or $(p_2+p_3)^2$ where
$p_3=p_B-p_1-p_2$.

Angular distributions of the baryon in the dibaryon rest frame
have also been measured in $p\bar p K^-,p\bar p K_S$ and
$\Lambda_c^+\bar p\pi^-$ modes. The measurements of the
correlation of the meson with the baryon provide the information
on the form-factor momentum dependence. For $\ov B\to \B_1\ov
\B_2M$, one can define the angular asymmetry as
 \be \label{eq:asy}
 A={N_+-N_-\over N_++N_-},
 \en
where $N_+$ and $N_-$ are the events with $\cos\theta>0$ and
$\cos\theta<0$, respectively, with $\theta$ being the angle
between the meson direction and the antibaryon $\ov \B_2$
direction in the dibaryon rest frame. The angular asymmetry is
measured to be $-0.59^{+0.08}_{-0.07}$, $0.32\pm0.14$, and
$0.36^{+0.23}_{-0.20}$ for $B^-\to p\bar p K^-, \Lambda_c^+\bar
p\pi^-,\Lambda\bar p\gamma$, respectively. Experimental
measurements indicate that, in the baryon-antibaryon rest frame,
the outgoing meson or the photon tends to emerge parallel to the
antibaryon for $B^-\to\Lambda_c^+\bar p\pi^-,\Lambda\bar p\gamma$
and to the baryon for $B^-\to p\bar p K^-$. As we shall see, a
stronger correlation of the meson to the antibaryon than to the
baryon in the dibaryon rest frame is expected in the pQCD picture
for $B$ decay. Hence, the opposite correlation effect seen in
$B^-\to p\bar p K^-$ by BarBar and Belle to be discussed below is
astonishing and entirely unexpected.

In the absence of theoretical guidance for the form factors in the
three-body matrix element (\ref{eq:3bodyme}), one may consider a
phenomenological pole model at the hadron level as put forward in
\cite{CYcharmless}. The main assumption of the pole model is that
the dominant contributions arise from the low-lying baryon and
meson intermediate states. As we shall see, the meson pole
diagrams are usually related to the vacuum to $\B_1\bar\B_2$
transition form factors and hence responsible for threshold
enhancement, whereas the baryon pole diagrams account for the
correlation of the outgoing meson with the baryon.

Consider a typical meson pole diagram, say, Fig.
\ref{fig:Qdiagrams}(b). The corresponding rate is proportional to
 \be
 \Gamma\propto \int\cdots\left|{g_{M'\B_1\ov \B_2}\over
 m_{12}^2-m_{M'}^2}\right|^2dm_{12}^2dm_{23}^2,
 \en
where $m_{ij}^2=(p_i+p_j)^2$ with $p_3=p_M$ and $M'$ is the mass
of the meson pole. It is clear that the meson propagator is
maximal when the dibaryon invariant mass is near threshold, i.e.
$m_{12}\sim m_1+m_2$. In general, the strong coupling at the
$M'\B_1\ov \B_2$ vertex can be related to the vacuum to
$\B_1\ov\B_2$ form factors via
 \be
 g_{M'\B_1\ov \B_2}(t)={t-m_{M'}^2\over f_{M'}m_{M'}}
 f^{\B_1\ov\B_2}(t)
 \en
with $t=m_{12}^2$. Then the current-induced decay amplitude is
reproduced and threshold enhancement is connected to the
asymptotic behavior of the form factors.

Next consider a baryon pole diagram such as the one in Fig.
\ref{fig:Qdiagrams}(a),
 \be
 \Gamma\propto \int\cdots\left|{g_{M\B_1\B_b}\over
 m_{13}^2-m_{\B_b}^2}\right|^2dm_{12}^2dm_{23}^2.
 \en
Since in the dibaryon rest frame
 \be
 m_{13}^2=m_1^2+m_M^2+2(E_1E_M-p_{c.m.}^2\cos\theta_{13}),
 \en
where $p_{c.m.}$ is the c.m. momentum and since $m_{\B_b}\gg
m_\B,m_M$, it is clear that the decay rate becomes prominent when
$\cos\theta_{13}\to -1$; that is, when the baryon is moving
antiparallel to the meson. Therefore, the baryon pole diagram
always implies that {\it the antibaryon tends to emerge parallel
to the outgoing meson}. Intuitively, the observation that the
meson is correlated more strongly to the $\ov \B_2$ than to the
$\B_1$ also can be understood in the following manner. Since in
the $B$ rest frame
 \be
 m_{12}^2=m_1^2+m_2^2+2(E_1 E_2-|\vec{p}_1||\vec{p}_2|\cos\theta_{12}),
 \en
threshold enhancement implies that the baryon pair $\B_1$ and $\ov
\B_2$ tend to move collinearly in this frame, i.e. $\theta_{12}\to
0$. From Fig. \ref{fig:Qdiagrams}(a) we see that the $\B_1$ is
moving faster than $\ov\B_2$ as the former picks up an energetic
quark from the $b$ decay.  When the system is boosted to the
$\B_1\bar\B_2$ rest frame, $\ov \B_2$ and $M$ are moving
collinearly away from the $\B_1$.

\subsection{$B^-\to p\bar p K^-$}

This mode has been measured by BaBar \cite{BaBar:ppK} and Belle
\cite{Belle:3charmless1} with the averaged branching ratio
$(6.10\pm0.48)\times 10^{-6}$ (Table \ref{tab:3charmless}).
Threshold enhancement in the dibaryon mass distribution is also
observed by both $B$ factories. Recently, Belle has studied the
angular distribution in the baryon-antibaryon pair rest frame
\cite{Belle:3charmless1}, while BaBar has measured the Dalitz plot
asymmetry in the decay $B^-\to p\bar pK^-$ \cite{BaBar:ppK}.

Based on the pole model and the intuitive argument described
before, the $K^-$ in the $p\bar p$ rest frame is expected to
emerge parallel to $\bar p$. However, the Belle observation is
other around \cite{Belle:3charmless1}: the $K^-$ is preferred to
move collinearly with the proton in the $p\bar p$ rest frame.
Instead of measuring angular distributions, BaBar has studied the
Dalitz plot asymmetry in the invariant masses $m_{pK}$ and
$m_{\bar pK}$ and found that $m_{pK}<m_{\bar pK}$. This is
consistent with the Belle result because in the $p\bar p$ rest
frame
 \be
 m_{pK}^2 &=& m_p^2+m_K^2+2(E_pE_K-|\vec{p}_K|p_{cm}\cos\theta_p), \non
 \\  m_{\bar pK}^2  &=& m_{\bar p}^2+m_K^2+2(E_{\bar p}E_K+|
 \vec{p}_K|p_{cm}\cos\theta_p),
 \en
where $\theta_p$ is the angle between the $K^-$ and the $p$
directions. Hence, the observation of $m_{pK}<m_{\bar pK}$ implies
that $\theta_p$ is preferred to be small. Therefore, $K^-$ and $p$
tend to move collinearly in the $p\bar p$ rest frame.

\begin{figure}[t]
\vspace{0cm}
\hspace{0cm}\centerline{\epsfig{figure=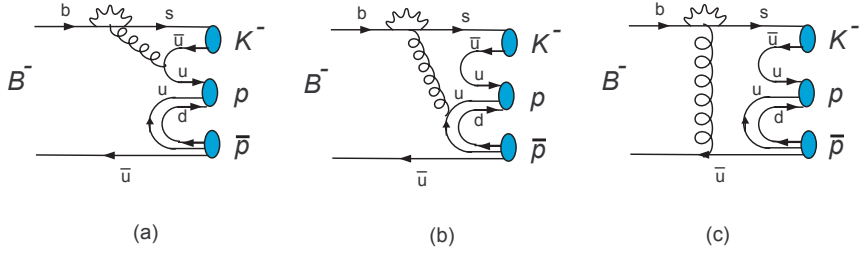,width=12cm} }
\vspace{-0.5cm}
    \caption{{\small Penguin diagram contributions to $B^-\to p\bar pK^-$.
    }} \label{fig:BppK}
\end{figure}

The correlation of the kaon with the proton observed by BaBar and
Belle is in contradiction to the naive expectation. Near the
threshold area, the proton and the antiproton move collinearly in
the $B$ rest frame. Again, $m_{pK}<m_{\bar pK}$ indicates that the
$p$ is moving slower than the $\bar p$. However, the dominant
factorizable penguin diagram for $B^-\to p\bar pK^-$ (Fig.
\ref{fig:BppK}(a)) implies a $p$ moving faster than $\bar p$. It
has been argued in \cite{Rosner} that since the $\bar u$ quark in
the $K^-$ is associated with a $u$ quark in the $p$ (Fig.
\ref{fig:BppK}), it leads to a strong $p-K^-$ angular correlation
and $m_{pK}<m_{\bar pK}$. However, this argument is valid only for
Fig. \ref{fig:BppK}(c) where $\bar p$ is moving faster than $p$.
For the dominant penguin diagram \ref{fig:BppK}(a), pole model or
the intuitive argument leads to an opposite correlation beween the
kaon and the proton. Since the penguin annihilation diagram Fig.
\ref{fig:BppK}(c) cannot be the dominant contribution to $B^-\to
p\bar pK^-$ as it is suppressed relative to the factorizable
penguin diagram by $1/m_B$, the observed angular distribution
becomes quite tantalizing.

The aforementioned puzzle could indicate that the $p\bar p$ system
is produced from some intermediate states, such as the glueball
and the baryonium, a $p\bar p$ bound state. This may change the
correlation pattern. This possibility is currently under study in
\cite{CCHT}. It is likely that the new mechanism responsible for
the anomalous correlation effect observed in $p\bar pK^-$ only
occurs in the penguin-dominated modes, but not in the
tree-dominated decay such as $B^-\to p\bar p\pi^-$. Hence,
experimentally it is very important to study the angular
distributions of the baryon for $p\bar p X$ with
$X=K^{*+},K^0,K^{*0},\pi^-$ and $\Lambda\bar\Lambda K^-$ modes.

\subsection{Other salient features}

\begin{figure}[t]
\includegraphics[width=4cm]{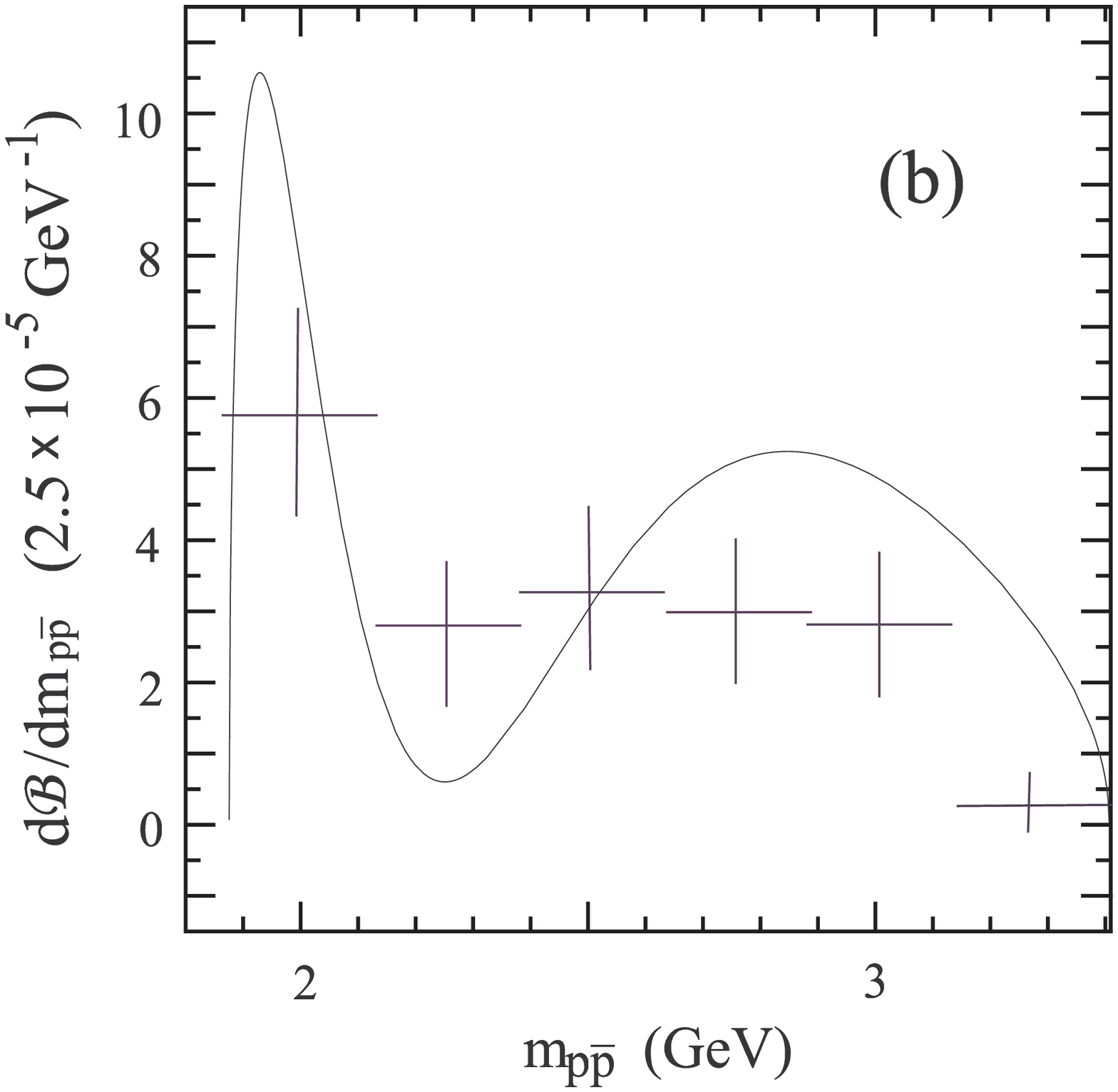}
\caption{The predicted $p\bar p$ invariant mass distribution of
$\ov B^0\to D^0p\bar p$ \cite{CKDmeson}. The experimental data are
taken from \cite{Belle:3charm}.} \label{fig:BDppinv}
\end{figure}

Many of other 3-body baryonic $B$ decays have been studied in
\cite{CKcharm,CYcharmless,CKDmeson,CHT01,CHT02,CH03,GHLamc}.
Because of space limitation, we will focus on a few of the
prominent features of them:

\begin{itemize}
\item Contrary to $\ov B^0\to D^{(*)+}n\bar p$ decays where the
$D^{*+}/D^+$ production ratio is anticipated to be of order 3, the
$D^{*0}/D^0$ production ratio in color suppressed $\ov B^0\to
D^{(*)0}p\bar p$ decays is consistent with unity experimentally
(see Table \ref{tab:3charm}). It is shown in \cite{CKDmeson} that
the similar rates for $D^0p\bar p$ and $D^{*0}p\bar p$ can be
understood within the framework of the pole model as the former is
dominated by the axial-vector meson states, whereas the other
modes proceed mainly through the vector meson poles.
 \item
The spectrum of $\ov B\to D^0p\bar p$ is predicted to have a hump
at large $p\bar p$ invariant mass $m_{p\bar p}\sim 2.9$ GeV
\cite{CKDmeson} (see Fig. \ref{fig:BDppinv}), which needs to be
checked by forthcoming experiments. As for the correlation, it is
expected that $\la m_{D\bar p}\ra<\la m_{Dp}\ra$ according to the
discussion shown in Sec.3.B, whereas it is other way around,
namely, $\la m_{Dp}\ra<\la m_{D\bar p}\ra$ in the fragmentation
picture of \cite{Rosner}.
 \item Charmless decays $B^-\to p\bar p
K^-(K^{*-})$ are penguin-dominated. It is naively expected that
$p\bar p K^{*-}<p\bar p K^-$ due to the absence of $a_6$ and $a_8$
penguin terms contributing to the former. The Belle observation of
a large rate for the $K^*$ production (see Table
\ref{tab:3charmless}) is thus unexpected. Also, it is non-trivial
to understand the observed sizable rate of $\ov B^0\to p\bar p \ov
K^0$ \cite{CYcharmless,CH03}.
 \item The relations among several penguin-dominated three-body
 baryonic decays
 \be
 && \Gamma(\ov B^0\to\Sigma^-\bar n\pi^+)= 2\Gamma(\ov B^0\to\Sigma^0\bar
p\pi^+)=2\Gamma(B^-\to\Sigma^-\bar
n\pi^0)=4\Gamma(B^-\to\Sigma^0\bar p\pi^0), \non \\
 && \Gamma(\ov B^0\to\Xi^0\bar\Sigma^-\pi^+)=2\Gamma(B^-\to\Xi^0
 \bar\Sigma^-\pi^0)=2\Gamma(\ov
 B^0\to\Xi^-\bar\Sigma^0\pi^+)=4\Gamma(B^-\to\Xi^-\bar\Sigma^0\pi^0), \non \\
 && \Gamma(\ov B^0\to\Lambda\bar p\pi^+)=2\Gamma(
 B^-\to\Lambda\bar p\pi^0),
 \en
have been derived based on isospin symmetry and factorization
\cite{CH03}. Hence, the factorization assumption can be tested by
measuring the above relations.
 \item The three-body doubly charmed baryonic decay
$B\to\Lambda_c\bar\Lambda_cK$ has been observed recently by Belle
with the branching ratio of order $7\times 10^{-4}$ (see Table
\ref{tab:3charm}). Since this mode is color-suppressed and its
phase space is highly suppressed, the naive estimate of $\B\sim
10^{-8}$ is too small by four to five orders of magnitude compared
to experiment. It was originally conjectured in \cite{CCT} that
the great suppression for the $\Lambda_c^+\bar\Lambda_c^-K$
production can be alleviated provided that there exists a narrow
hidden charm bound state with a mass near the
$\Lambda_c\bar\Lambda_c$ threshold. This possibility is plausible,
recalling that many new charmonium-like resonances with masses
around 4 GeV starting with $X(3872)$ \cite{X3872} and so far
ending with $Y(4260)$ \cite{Y4260} have been recently observed by
BaBar and Belle. This new state that couples strongly to the
charmed baryon pair can be searched for in $B$ decays and in
$p\bar p$ collisions by studying the mass spectrum of $D^{(*)}\ov
D^{(*)}$ or $\Lambda_c\bar\Lambda_c$. However, no new resonance
with a mass near the $\Lambda_c\bar\Lambda_c$ threshold was found
(see Fig. 3 in version 2 of \cite{Belle:2LamcK}). This implies the
failure of naive factorization for this decay mode and may hint at
the importance of nonfactorizable contributions such as
final-state effects. For example, the weak decay $B\to D^{(*)}\bar
D_s^{(*)}$ followed by the rescattering $D^{(*)}\bar D_s^{(*)}\to
\Lambda_c\bar\Lambda_c K$ or the decay $B\to \Xi_c\bar\Lambda_c$
followed by $\Xi_c\bar\Lambda_c\to\Lambda_c\bar\Lambda_cK$ may
explain the large rate observed for $B\to\Lambda_c\bar\Lambda_cK$.

\item Triple product correlations (TCP) in three-body baryonic $B$
decays can be studied to test $T$ violation. Unlike the usual
direct CP asymmetry, the $T$-odd asymmetry due to TCP does not
vanish even in the absence of strong phases. It has been estimated
in \cite{GHTodd} that $T$ violation induced  from the asymmetry in
$\vec{s}_\Lambda\cdot(\vec{p}_{\bar p}\times \vec{p}_\Lambda)$ in
$\ov B^0\to\Lambda\bar p\pi^+$ decay can be as large as 10\%,
while CP asymmetry is only at 1\% level.

\end{itemize}

\section{Radiative baryonic $B$ decays}
Naively it appears that the bremsstrahlung process will lead to
$\Gamma(B\to\B_1\ov \B_2\gamma)\sim {\cal O}(\alpha_{\rm
em})\Gamma(B\to\B_1\ov \B_2)$ with $\alpha_{\rm em}$ being an
electromagnetic fine-structure constant and hence the radiative
baryonic $B$ decay is further suppressed than the two-body
counterpart, making its observation very difficult at the present
level of sensitivity for $B$ factories. However, there is an
important short-distance electromagnetic penguin transition $b\to
s \gamma$. Owing to the large top quark mass, the amplitude of
$b\to s\gamma$ is neither quark mixing nor loop suppressed.
Moreover, it is largely enhanced by QCD corrections. As a
consequence, the short-distance contribution due to the
electromagnetic penguin diagram dominates over the bremsstrahlung.
This phenomenon is quite unique to the bottom hadrons which
contain a heavy $b$ quark; such a magic short-distance enhancement
does not occur in the systems of charmed and strange hadrons.

\begin{figure}[t]
\vspace{-1cm}
\hspace{0cm}\centerline{\psfig{figure=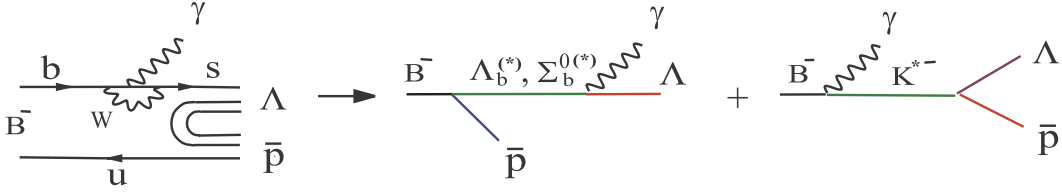,width=15cm}}
\vspace{-1cm}
    \caption{{\small Quark and pole diagrams for $B^-\to\Lambda\bar
    p\gamma$.
    }}
    \label{fig:Lampgammadia}
\end{figure}

Since a direct evaluation of this radiative decay is difficult as
it involves an unknown 3-body matrix element
$M_{\mu\nu}=\la\Lambda\bar p|\bar
s\sigma_{\mu\nu}(1+\gamma_5)b|B^-\ra$, we shall instead evaluate
the corresponding diagrams known as pole diagrams at the hadron
level (see Fig. \ref{fig:Lampgammadia}). In principle, there exist
many possible baryon and meson pole contributions. The main
assumption of the pole model is that the dominant contributions
arise from the low-lying baryon and meson intermediate states. For
$B^-\to\Lambda\bar p\gamma$, the relevant intermediate states are
$\Lambda_b^{(*)}$, $\Sigma_b^{0(*)}$ and $K^*$.

The predicted branching ratios for $B^-\to\Sigma^0\bar p\gamma,~
\Xi^0\bar\Sigma^-\gamma$ and $\Xi^-\bar\Lambda\gamma$  decays are
summarized in Table \ref{tab:radBR}.\footnote{In the previous work
\cite{CYrad}, the meson pole contributions which are {\it a
priori} not necessarily small have been neglected. It is updated
in \cite{CYrad06} including some improved input parameters such as
strong couplings.}
Decay rates for the other modes can be obtained via the relations
\cite{CYrad} (see also \cite{GHrad,Kohara}
 \be
 && \Gamma(B^-\to\Sigma^-\bar n\gamma)= 2\Gamma(B^-\to\Sigma^0\bar
p\gamma)=2\Gamma(\ov B^0\to\Sigma^0\bar
n\gamma)=\Gamma(\ov B^0\to\Sigma^+\bar p\gamma), \non \\
 && \Gamma(\ov B^0\to\Xi^-\bar\Sigma^+\gamma)=2\Gamma(B^-\to\Xi^-
 \bar\Sigma^0\gamma)=2\Gamma(\ov
 B^0\to\Xi^0\bar\Sigma^0\gamma)=\Gamma(B^-\to\Xi^0\bar\Sigma^-\gamma), \non \\
 && \Gamma(\ov B^0\to\Lambda\bar n\gamma)=\Gamma(
 B^-\to\Lambda\bar p\gamma), \qquad \Gamma(\ov
 B^0\to\Xi^0\bar\Lambda\gamma)=\Gamma(B^-\to\Xi^-\bar\Lambda\gamma).
 \en
It is interesting to notice that the $\Sigma^0\bar p\gamma$ mode,
which was previously argued to be very suppressed due to the
smallness of the strong coupling $g_{\Sigma_b\to B^-p}$
\cite{CYrad}, receives the dominant contribution from the $K^*$
pole diagram and its branching ratio is consistent with that
obtained in \cite{GHrad}. In contrast, the mode
$\Xi^0\bar\Sigma^-\gamma$ is dominated by the baryon pole
contribution. Meson and baryon intermediate state contributions
are comparable in $\Lambda\bar p\gamma$ and
$\Xi^-\bar\Lambda\gamma$ modes except that they interfere
constructively in the former but destructively in the latter.
Recently, Belle \cite{Belle:Lampgam} has made the first
observation of radiative hyperonic $B$ decay $B^-\to\Lambda\bar
p\gamma$ with the result
 \be
 \B(B^-\to\Lambda\bar p\gamma)=(2.16^{+0.58}_{-0.53}\pm0.20)\times
 10^{-6}.
 \en
In addition to the first observation of $\Lambda\bar p\gamma$, the
decay $B^-\to \Xi^0\bar\Sigma^-\gamma$ at the level of $6\times
10^{-7}$ may be accessible to $B$ factories in the future.

\begin{table}[t]
\caption{Branching ratios and angular asymmetries defined in Eq.
(\ref{eq:asy}) for radiative baryonic $B$ decays.}
\label{tab:radBR}
\begin{ruledtabular}
\begin{tabular}{l l l l c }
 Mode & Baryon pole & Meson pole & Br(total) & Angular asymmetry \\ \hline
$B^-\to\Lambda\bar p\gamma$ & $7.9\times 10^{-7}$ & $9.5\times
10^{-7}$ & $2.6\times 10^{-6}$ & 0.25 \\
$B^-\to\Sigma^0\bar p\gamma$ & $4.6\times 10^{-9}$ &
$2.5\times 10^{-7}$ & $2.9\times 10^{-7}$ & 0.07 \\
$B^-\to\Xi^0\bar\Sigma^-\gamma$ & $7.5\times 10^{-7}$ & $1.6\times
10^{-7}$ & $5.6\times 10^{-7}$ & 0.43 \\
$B^-\to\Xi^-\bar\Lambda\gamma$ & $1.6\times 10^{-7}$ & $2.4\times
10^{-7}$ & $2.2\times 10^{-7}$ & 0.13 \\
\end{tabular}
\end{ruledtabular}
\end{table}

\begin{figure}[t]
\vspace{0cm} \centerline{
            {\epsfxsize2.8in \epsffile{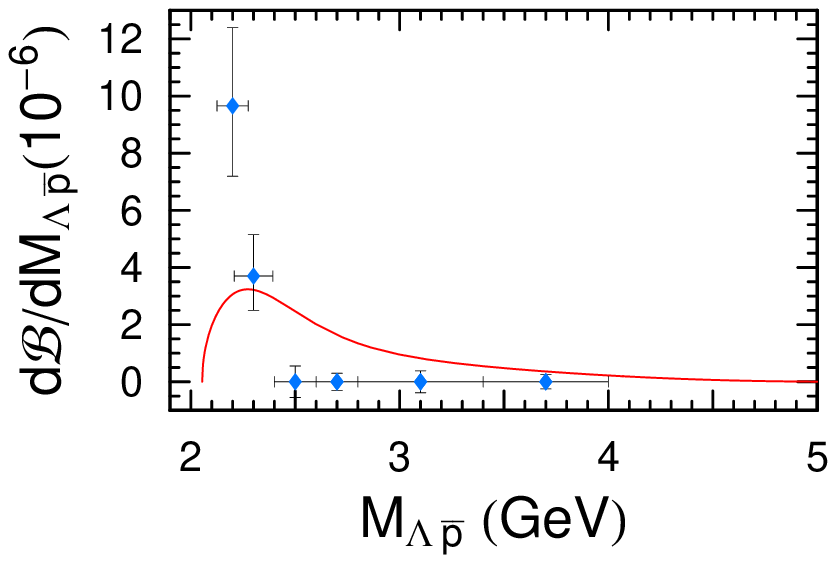}}
           {\epsfxsize2.65in \epsffile{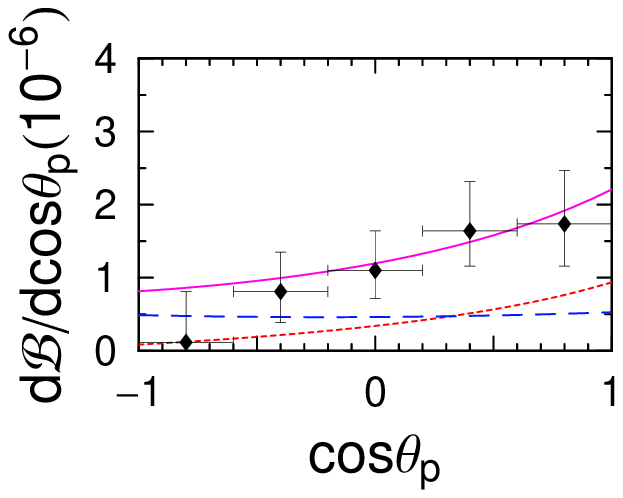}}}
 \centerline{\,\,\,\,\,(a)\hspace{7.2cm}(b)} \vskip0.2cm
\caption{(a) $\Lambda\bar p$ invariant mass distribution and (b)
the angular distribution of the antiproton in the baryon pair
system for $B^-\to\Lambda\bar p\gamma$, where $\theta_p$ is the
angle between the antiproton direction and the photon direction in
the $\Lambda \bar p$ rest frame. The dotted, dashed and solid
curves stand for baryon pole, meson pole and total contributions,
respectively. Data are taken from \cite{Belle:Lampgam}.}
\label{fig:Lampgamma}
\end{figure}

In addition to the threshold enhancement effect observed in the
differential branching fraction of $\Lambda \bar p\gamma$ (Fig.
\ref{fig:Lampgamma}.(a)), Belle has also measured the angular
distribution of the antiproton in the $\Lambda\bar p$ system (Fig.
\ref{fig:Lampgamma}.(b)), where $\theta_p$ is the angle between
the antiproton direction and the photon direction in the $\Lambda
\bar p$ rest frame. It is clear that the $\Lambda$ tends to emerge
opposite the direction of the photon. The angular asymmetry is
measured by Belle to be $A=0.36^{+0.23}_{-0.20}$ for
$B^-\to\Lambda\bar p\gamma$ \cite{Belle:Lampgam}. The meson pole
diagram is responsible for low-mass enhancement and does not show
a preference for the correlation between the baryon pair and the
photon (see the dashed curve in Fig. \ref{fig:Lampgamma}). Our
prediction $A=0.25$ (see Table \ref{tab:radBR}) is consistent with
experiment.

\section{Conclusion}
Experimental and theoretical progresses in exclusive baryonic $B$
decays in the past five years are impressive. The threshold
peaking effect in baryon pair invariant mass is one of the key
ingredients in understanding three-body decays. Weak radiative
baryonic decays  mediated by the electromagnetic penguin process
$b\to s\gamma$ are studied and some of them are readily accessible
experimentally. There are two unsolved puzzles with the 3-body
decays: one is the anomalous correlation effect observed in
$B^-\to p\bar pK^-$ decay and the other is the unexpectedly large
rate observed for $B\to\Lambda_c\bar\Lambda_cK$. The former may
indicate that $p\bar p$ is coupled to some intermediate states,
while the latter implies the failure of naive factorization for
$\Lambda\bar\Lambda K$ modes and may hint at the importance of
final-state rescattering effects. Experimentally, it is very
important to measure the correlation of the outgoing meson with
the baryon in three-body baryonic $B$ decays in order to gain
further dynamical insight and to discriminate between different
models.

\begin{acknowledgments}
I'm grateful to Chun-Khiang Chua, Shang-Yuu Tsai and Kwei-Chou
Yang for fruitful collaboration, to Min-ru Wang for valuable
discussions and to Otto Kong for organizing this excellent
conference.

\end{acknowledgments}

%%%%%%%%%%%%%%%%%%%%%%%%%%%%%%%%%%%%%%%%%%%%%%%%%%%%%%%%

\end{document}